\documentclass[namedreferences]{solarphysics}
\usepackage[hyperref,optionalrh,solaromanenum]{spr-sola-addons} 
\usepackage{graphicx}                    
\usepackage{color}                       
\usepackage{breakurl}                         

\renewcommand{\vec}[1]{{\mathbfit #1}}
\newcommand{\deriv}[2]{\frac{{\mathrm d} #1}{{\mathrm d} #2}}
\newcommand{\secderiv}[2]{\frac{{\mathrm d}^2 #1}{{\mathrm d} {#2}^2}}

\newcommand{\rmd}{ {\ \mathrm d} }
\newcommand{\uvec}[1]{ \hat{\mathbf #1} }
\newcommand{\pder}[2]{ \frac{\partial #1}{\partial #2} }
\newcommand{\psecder}[2]{ \frac{\partial^2 #1}{\partial {#2}^2} }

\newcommand{\grad}{ {\bf \nabla } }
\newcommand{\curl}{ {\bf \nabla} \times}

\newcommand{\bb}{\vec B}
\newcommand{\jj}{ \vec j}

\newcommand{\hypf}{{~}_2{\mathrm F}_1}

\newcommand{\aap}{    {\it Astron. Astrophys.}}

\newcommand{\aapr}{   {\it Astron. Astrophys. Rev.}}

\newcommand{\apj}{    {\it Astrophys. J.}}

\newcommand{\apjs}{   {\it Astrophys. J. Suppl. Ser.}}

\newcommand{\gafd}{   {\it Geophys. Astrophys. Fluid Dyn.}}

\newcommand{\jgr}{    {\it J. Geophys. Res.}}
\newcommand{\mnras}{  {\it Mon. Not. Roy. Astron. Soc.}}

\newcommand{\solphys}{{\it Solar Phys.}}
 
\newcommand{\ssr}{    {\it Space Sci. Rev.}}

\begin{document}

\begin{article}

\begin{opening}

\title{Analytical Three-dimensional Magnetohydrostatic Equilibrium Solutions 
for Magnetic Field Extrapolation 
Allowing a Transition from Non-force-free to Force-free Magnetic Fields}

%
\author[addressref=aff1,corref,email={tn3@st-andrews.ac.uk}]{\inits{T.}\fnm{Thomas}~\lnm{Neukirch}\orcid{0000-0002-7597-4980}}
\author[addressref=aff2,email={wiegelmann@mps.mpg.de}]{\inits{T.}\fnm{Thomas}~\lnm{Wiegelmann}}

%

%
  \address[id=aff1]{School of Mathematics and Statistics, 
                            University of St Andrews, 
                            St Andrews, 
                            KY16 9SS, 
                            United Kingdom}
  \address[id=aff2]{Max-Planck-Institut f\"ur Sonnensystemforschung,
                            Justus-von-Liebig-Weg 3,
                            37077 G\"ottingen, 
                            Germany}

\runningauthor{T.\ Neukirch, T.\ Wiegelmann}
\runningtitle{3D MHS Equilibria for Extrapolation}

\begin{abstract}
For the extrapolation of magnetic fields into the solar corona from measurements taken in the photosphere (or chromosphere) 
force-free magnetic fields are typically used. This does not take into account that the lower layers of the solar atmosphere are not force-free.
While some numerical extrapolation methods using magnetohydrostatic magnetic fields have been suggested, a complementary and numerically comparatively cheap method
is to use analytical magnetohydrostatic equilibria to extrapolate the magnetic field.
In this paper, 
we present a new family of solutions for a special class of analytical three-dimensional magnetohydrostatic equilibria, which can
be of 
use 
for such magnetic field extrapolation. 
The new solutions 
allow for the more flexible
modelling of a transition from non-force-free to (linear) force-free magnetic fields. 
In particular,
the height and width of the region where this transition takes place can be specified by choosing appropriate model parameters.
\end{abstract}

%

\end{opening}

%
 \section{Introduction}%
 \label{sec:introduction} 
 
Modelling the magnetic field in the solar atmosphere is of great importance for our interpretation of many of solar observations, in particular in the solar corona \citep[\textit{e.g.}][]{wiegelmann:etal14}. 
Because coronal magnetic fields cannot be measured routinely with the required accuracy, extrapolation methods with photospheric magnetic field measurements as boundary 
conditions are normally used, usually assuming that the magnetic field is force-free
\citep[see \textit{e.g.} recent reviews by][]{wiegelmann:sakurai12, regnier13}. In recent years measurements of the magnetic field in the chromosphere have also become 
available \citep[\textit{e.g.}][]{harvey12}. An overview of measurements of photospheric and chromospheric magnetic fields can, for example, be found in the paper by \citet{lagg:etal17}.

While the assumption of force-free magnetic fields is well satisfied in large parts of the solar corona due to the low plasma $\beta$, the lower parts of the solar 
atmosphere (chromosphere and photosphere) can in general not be considered to be force-free  
\citep[\textit{e.g.}][]{metcalf:etal95,De_Rosa:etal09,wiegelmann:etal14} and hence magnetohydrostatic (MHS) models, including pressure and gravity, would be more appropriate for these regions. 
Developing numerical extrapolation methods for the magnetostatic case has, for example, been attempted 
in \citet{wiegelmann:neukirch06} (including pressure only), \citet{gilchrist:wheatland13}, and \citet{zhu:wiegelmann18}.

These numerical approaches are usually computationally expensive. Hence, a complementary
approach 
for including pressure and gravity forces in magnetic extrapolation,
which is computationally relatively cheap, would be to use analytical three-dimensional MHS equilibrium solutions. Obviously, just as for 3D force-free fields 
only a limited number of analytical 3D MHS solutions suitable for magnetic field extrapolation are known, with the known 3D MHS solutions 
useful for extrapolation purposed being comparable in status to 3D linear force-free solutions. 
We emphasize that in order to find analytical solutions to the MHS equations in 3D one has to make a number of assumptions, which may limit the applicability of the method to some extent. Hence, this approach using analytical MHS equilibria has
to be regarded as an alternative method which allows one to get a reasonably fast extrapolation method including a non-force-free part of the solar atmosphere, but not as a replacement for the
numerical approaches mentioned above.

Various aspects of the theory of analytical 3D MHS solutions have been developed in a series of papers 
by \citet{low82, low84, low85, low91, low92, low93a, low93b, low05} and \citet{bogdan:low86}, both in Cartesian\footnote{In this paper, we designate as "Cartesian" solutions all 
solutions that use a constant gravitational force along one Cartesian direction; this includes solutions that are actually formulated in cylindrical polar coordinates.} and in spherical geometry.
Additions, extensions and applications of this work were provided by, for example, \citet{neukirch95a,neukirch97c,neukirch97b}, \citet{neukirch:rastatter99}, \citet{petrie:neukirch00}, \citet{neukirch09}, \citet{al-salti:etal10}, \citet{al-salti:neukirch10}, \citet{gent:etal13}, \cite{gent:etal14}, \citet{mactaggart:etal16} and \citet{wilson:neukirch18}. A different, but less general approach has been pursued by \citet{osherovich85a, osherovich85b}. 

Subsets of these three-dimensional MHS solutions have been used for modelling both global solar magnetic field models, using spherical coordinates 
\citep[\textit{e.g.}][]{bagenal:gibson91, gibson:bagenal95, gibson:etal96, zhao:hoeksema93, zhao:hoeksema94, zhao:etal00, ruan:etal08}, and local solar coronal structures, using Cartesian or cylindrical coordinates
\citep[\textit{e.g.}][]{aulanier:etal99, aulanier:etal98b, petrie-phd00, gent:etal13, gent:etal14,wiegelmann:etal15,wiegelmann:etal17}. Other applications include, for example, models of the magnetic fields of stars \citep[\textit{e.g.}][]{mactaggart:etal16} and their interaction with 
exoplanets \citep[\textit{e.g.}][]{lanza08,lanza09}. In this paper, we shall focus on 3D MHS solutions in Cartesian geometry, {\it i.e.} assuming a constant gravitational force, which we take to point in the negative $z$-direction (hence the coordinate $z$ has the meaning of height above the photosphere from now on).

If we follow the argument that small plasma-$\beta$ should imply nearly force-free magnetic fields (and vice versa), there should be a marked transition
from non-force-free to force-free fields with increasing height when moving from the photosphere through the chromosphere into the corona. Of the currently known solutions the one which comes closest to 
showing this feature has been suggested by \citet{low91,low92} and has an exponential height profile of the non-force-free current density. This solution has been used repeatedly for modelling purposes 
\citep[see \textit{e.g.}][]{aulanier:etal98b, aulanier:etal99,wiegelmann:etal15, wiegelmann:etal17}. It is the aim of this paper to provide another set of 3D MHS solutions whose non-force-free current density have a more flexible dependence on height.

A general problem which arises in the use of this class of 3D MHS solutions for 
extrapolation
purposes is that one has to be careful to avoid generating regions of negative plasma pressure or density 
\citep[see \textit{e.g.}][]{petrie:neukirch00,petrie-phd00,gent:etal13}.
This problem is caused by the mathematical structure of the expressions for the plasma pressure and density, both of which are written as the difference between a positive background pressure and density, and 
potentially negative terms, depending on the {\em a priori} unknown magnetic field solution. In theory, this problem can always be solved by either increasing the background plasma pressure and density or by decreasing the amplitude of the negative terms. In practice, the former often leads to an unrealistically high value of the plasma-$\beta$ throughout the model domain, whereas the latter can cause the loss
of meaningful spatial structures in plasma pressure and density. Being able to control the solution structure better should 
be of advantage for modelling purposes and may also have a (positive) bearing on the problem of keeping plasma pressure and density positive everywhere.

The structure of the paper is as follows. In Section \ref{sec:theory} we briefly summarise the basic theory of the particular class of 3D MHS solutions we use in this paper and 
then present the calculation
leading to the 
the new 
set of 
solutions in Section \ref{sec:solutions}.
In Section \ref{sec:examples}, we present some example solutions and in Section \ref{sec:discussion} a discussion and conclusions.

\section{Basic Theory}
\label{sec:theory}

The MHS equations are given by
\begin{eqnarray}
\jj \times \bb - \grad p - \rho \grad \Psi &=& 0 , \label{eq:forcebalance} \\
\curl \bb & = & \mu_0 \jj      ,                                    \label{eq:ampere} \\
\nabla \cdot \bb & =& 0           .                                             \label{eq:solenoidal}
\end{eqnarray}
Here $\bb$ denotes the magnetic field, $\jj$ the current density, $p$ the plasma pressure, $\rho$ the mass density, $\Psi$ the gravitational potential and $\mu_0$ is the permeability of free space.

In this paper we use Cartesian geometry with a constant gravitational force pointing in the negative
$z$-direction, {\it i.e.}\ $\Psi = g z$ with $g$ being the constant gravitational acceleration. 
The general theory for this case was first developed by \citet{low91,low92}. Later, \citet{neukirch:rastatter99} presented a somewhat simpler, albeit equivalent, formulation 
and we will follow their approach in the 
following brief
summary. The main assumption made is that the current density can be written in the form
\begin{equation}
\mu_0\jj = \alpha \bb + \curl (F \uvec{z}).
\label{eq:c-density-low91}
\end{equation}
It can be shown \citep{low91,neukirch:rastatter99} that $F$ has to be a function of $B_z$ and $z$. The form for $F$ suggested by \citet{low91,low92} was
\begin{equation}
F = f(z) B_z,
\label{eq:F-low91}
\end{equation}
{\it i.e.} linear in $B_z$, but with an arbitrary function $f(z)$. Clearly, the function $F$ is responsible for the non-force-free, {\it i.e.} perpendicular, part of the current density in this approach, although it should be noted that 
it will generally also contribute to the parallel part of the current density.
Hence, the choice of the free function $f(z)$ influences the dependence of the non-force-free part of the current density with height. 

Choosing $F$ to be a linear in $B_z$ leads to a linear partial differential equation
for $B_z$ \citep[\textit{e.g.}][]{low91}. Alternatively, representing the magnetic field $\bb$ in the form \citep[\textit{e.g.}][]{nakagawa:raadu72}
\begin{equation}
\bb = \curl[ \curl (\Phi \uvec{z}) ]+ \curl (\Theta\uvec{z}),
\label{eq:b-naka-raadu}
\end{equation}
one can show that \citep{neukirch:rastatter99}
\begin{equation}
\Delta \Phi - f(z) \Delta_{xy} \Phi + \alpha^2 \Phi = 0,
\label{eq:Phi-PDE}
\end{equation}
where
\begin{equation}
\Delta \Phi = \psecder{\Phi}{x} + \psecder{\Phi}{y} +\psecder{\Phi}{z} 
\label{eq:laplaceop}
\end{equation}
is the Laplace operator, and
\begin{equation}
\Delta_{xy} \Phi = \psecder{\Phi}{x} + \psecder{\Phi}{y} 
\label{eq:laplaceop2d}
\end{equation}
is the two-dimensional Laplace Operator in $x$ and $y$. Additionally, one finds that
\begin{equation}
\Theta=\alpha \Phi,
\label{eq:Theta-def}
\end{equation}
and hence that
\begin{eqnarray}
B_x &= & \frac{\partial^2 \Phi}{\partial x \partial z} + \alpha \pder{\Phi}{y}, \label{eq:bx-def} \\
B_y &= & \frac{\partial^2 \Phi }{\partial y \partial z} - \alpha \pder{\Phi }{x}, \label{eq:by-def} \\
B_z &=& - \psecder{\Phi}{x} - \psecder{\Phi}{y}.                                           \label{eq:bz-def}
\end{eqnarray}
The plasma pressure and the plasma density are given by the expressions
\begin{eqnarray}
p  &=& p_0(z) - f(z) \frac{B_z^2}{2\mu_0},  \label{eq:pressure-def} \\
\rho &=& \frac{1}{g}\left(-\deriv{p_0}{z} + \deriv{f}{z}  \frac{B_z^2}{2\mu_0}+ \frac{f}{\mu_0} \bb \cdot\nabla B_z  \right),
\label{eq:rho-def}
\end{eqnarray}
where $g$ is the constant gravitational acceleration. The plasma temperature can be determine from the plasma density and pressure {\it via}  an equation of state, for example, the ideal
gas law
\begin{equation}
T = \frac{\bar{\mu} p}{k_{\mathrm{B}} \rho},
\label{eq:idealgaslaw}
\end{equation}
with $\bar{\mu}$ the mean atomic weight of the plasma and $k_{\mathrm{B}}$ the Boltzmann constant.

\section{A New Family of Solutions}
\label{sec:solutions}

Solutions to Equation \ref{eq:Phi-PDE} have been found for several different choices of the function $f(z)$, using either separation of variables \citep[see \textit{e.g.}][]{low91} or the Green's function method \citep[\textit{e.g.}][]{petrie:neukirch00}. We will use separation of variables in this paper. Leaving $f(z)$ unspecified for the time being, separation of variables leads to 
\begin{equation}
\Phi= \int\!\!\! \!\int\limits_{\mbox{\hspace{-12pt}}-\infty}^{\mbox{\hspace{-4pt}}\infty} \bar{\Phi}(z; k_x, k_y) \exp[ \mbox{i}(k_x x + k_y y)] \rmd k_x\!\! \rmd k_y
\label{eq:Phi-cartesian}
\end{equation}
in Cartesian coordinates $x$, $y$, $z$ 
with the equation for $\bar{\Phi}$ being
\begin{equation}
\secderiv{\bar{\Phi}}{z} + [\alpha^2 -k^2+ k^2 f(z)] \bar{\Phi} =0,
\label{eq:Phibarequation}
\end{equation}
where $k^2 = k_x^2 +k_y^2$.
In Equation \ref{eq:Phibarequation} we have suppressed the parametric dependence of $\bar{\Phi}$ on $k_x$ and $k_y$ in the Cartesian
coordinate case 
(or on $k$ and $m$ in the cylindrical coordinate case). At this point it is also convenient to normalise 
all coordinates by a typical length scale $L$, as well as regarding $\alpha$ as the normalised $\alpha L$ and $k$ as $ k L$ from now on.

In practical applications, for example using potential or linear force-free magnetic fields to extrapolate photospheric field measurements, different versions of periodic boundary conditions are often imposed 
\citep[\textit{e.g.}][]{seehafer78,alissandrakis81,otto:etal07}.
If periodic boundary conditions in $x$ and $y$ are imposed $k_x$, $k_y$ take on discrete values and the integrals in Equation \ref{eq:Phi-cartesian} are replaced by infinite sums (which are then truncated in applications
to observational data). With such boundary conditions Fast Fourier Transformation (FFT) techniques can be used to implement the boundary conditions very effectively.

As pointed out by \citet{neukirch95a} Equation \ref{eq:Phibarequation} is analogous to a one-di\-men\-sional Schr\"odinger equation with the potential $-k^2 f(z)$ 
with energy eigenvalue $-(\alpha^2 -k^2)$ (or alternatively $-k^2[f(z) -1]$ with energy eigenvalue $-\alpha^2$). Thus solutions to Equation \ref{eq:Phibarequation}
are known for a large class of functions $f(z)$, but not all of these will be of use for modelling solar magnetic fields.
Examples of functions $f(z)$ which have been used for modelling solar magnetic fields are $f(z)=f_0=$constant, for which the solutions for $\bar{\Phi}$ are 
exponential functions \citep[\textit{e.g.}][]{neukirch97c,petrie-phd00, petrie:neukirch00}, and $f(z)= \bar{a} \exp(-\kappa z)$, with Bessel functions with an argument that is proportional to $ \exp(-\kappa z/2)$
as solutions \citep[\textit{e.g.}][]{low91, low92, aulanier:etal98b, aulanier:etal99}.

In this paper we aim to find solutions to Equations \ref{eq:Phi-PDE} or \ref{eq:Phibarequation}, respectively, which show a transition from non-force-free behaviour
to (linear) force-free behaviour as the height $z$ increases from the photosphere through the chromosphere into the corona. Hence, we would like $f(z)$ to 
possibly decrease from non-zero values
to a value close to zero when reaching the corona. While it is also in principle possible to model such a behaviour with the exponential $f(z)$ mentioned above,
the transition is only controlled by the parameter $\kappa$.
Because we would like to be able to control both the height at which the transition from non-force-free to force-free fields happens as well as the range in height
over which this happens we here suggest to use the function
\begin{equation}
f(z) = a\left[1 - b \tanh\left(\frac{z-z_0}{\Delta z} \right)\right],
\label{eq:eckart-f-def}
\end{equation}
with $a$, $b$, $z_0$ and $\Delta z$ real parameters.
The parameters $a $ and $b $ are dimensionless and we assume $a > 0$ and $1\ge b >0 $ in this paper. 
At $z=0$ we have
\begin{equation}
f(0) = a\left[1 +b \tanh\left(\frac{z_0}{\Delta z} \right)\right].
\label{eq:f-at-0}
\end{equation}

The parameters $a$ and $b$ control the overall magnitude of $f$ in the regimes $(z- z_0)/\Delta z \ll 0$ and $(z- z_0)/\Delta z \gg 0$. It is easy to see that
$f \to a(1+b)$ for $(z- z_0)/\Delta z \ll 0$ and 
$f \to a(1-b) $ for $(z- z_0)/\Delta z \gg 0$. The parameter $z_0$ controls where this transition takes place and the parameter $\Delta z$ is the length scale over which the transition happens. We see that $f \to 0$
for $(z- z_0)/\Delta z \gg 0$ if $b=1$ is chosen. This means that any perpendicular electric currents go to zero above the region in which the transition happens and the magnetic field becomes a (linear) force-free field.

 \begin{figure}    
   \centerline{\includegraphics[width=0.75\textwidth,clip=]{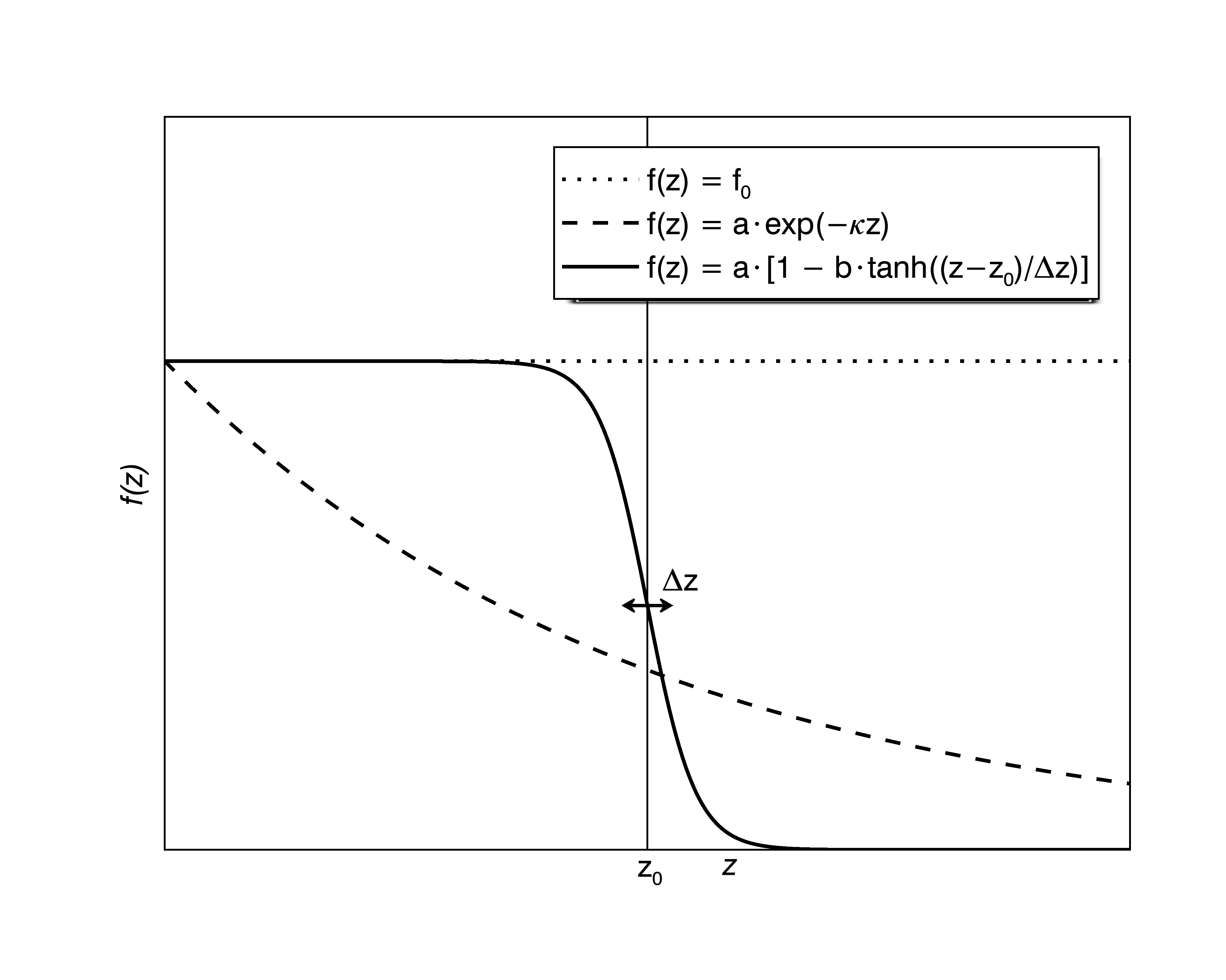}
              }
              \caption{The shape of $f(z)$ (solid line) defined in Equation \ref{eq:eckart-f-def}. Also shown are the two limiting cases described  in the main text.}
   \label{fig:fofz}
   \end{figure}

We remark that the form of $f(z)$ in Equation \ref{eq:eckart-f-def} includes the possibility of $f(z)= f_0 =\mbox{constant}$, by choosing $b=0$. As discussed above, for this special case the $z$-dependence of
the solutions is given by exponential functions \citep[\textit{e.g.}][]{neukirch97c,petrie-phd00, petrie:neukirch00}. In principle, Equation \ref{eq:eckart-f-def} also includes the case of an exponentially decaying function of $z$ 
\citep[\textit{e.g.}][]{low91}
by
choosing $b=1$ and letting 
\begin{equation}
\lim\limits_{z_0 \to -\infty} a \exp(2 z_0/\Delta z) = \bar{a} > 0.
\label{eq:exp-limit}
\end{equation}
As stated above,
in this case the $z$-dependence is given by Bessel functions with exponential arguments of the form $\exp(-\kappa z/2)$, with $\kappa/2 = 1/(\Delta z)$ in our limit \citep[see \textit{e.g.}][]{low91}. However,
as one can see from the expression inside the limit, this case combines two of the parameters of $f$ into one parameter $\bar{a}$ and hence has less flexibility in applications.
Although, as discussed before, the model with an exponential $f(z)$ also allows for a transition from non-force-free to force-free fields one can only control the length scale over which this happens, but one does not have an
additional specific height at which this happens such as $z_0$ in the $f$ considered in this paper. 
We show example plots of the three different forms of $f(z)$ discussed above in Figure\ \ref{fig:fofz}.

Substituting the function $f(z)$ given in Equation \ref{eq:eckart-f-def} into Equation \ref{eq:Phibarequation} we obtain:
\begin{equation}
\secderiv{\bar{\Phi}}{z} + \left[\alpha^2 -k^2(1-a) - k^2 a b \tanh\left(\frac{z-z_0}{\Delta z} \right)
 \right] \bar{\Phi} =0,
\label{eq:Phibar-eckart-equation}
\end{equation}
Making use of the coordinate transformation
\begin{equation}
\eta = \frac{1}{2}\left[1-   \tanh\left(\frac{z-z_0}{\Delta z} \right) \right]
\label{eq:eckart-transform}
\end{equation}
with $0 < \eta < 1$, we can transform Equation \ref{eq:Phibar-eckart-equation} into the differential equation
\begin{equation}
\secderiv{\bar{\Phi}}{\eta} +  \left( \frac{1}{\eta} - \frac{1}{1-\eta}\right)\deriv{\bar{\Phi}}{\eta} + 
 \left( \frac{C_1}{\eta} - \frac{C_2}{1-\eta}\right) \frac{\bar{\Phi}}{\eta (1-\eta)} = 0,
\label{eq:Phibar-hypergeometric-equation}
\end{equation}
where
\begin{eqnarray}
C_1 &=&  \frac{ 1}{4} \left[ {\bar{k}}^2 (1-a +ab) - {\bar{\alpha}}^2 \right],  \label{eq:C1} \\
C_2 &=&  \frac{ 1}{4} \left[ {\bar{k}}^2 (1-a -ab) - {\bar{\alpha}}^2 \right],  \label{eq:C2}
\end{eqnarray}
with $\bar{k} = k \Delta z$ and $\bar{\alpha} = \alpha \Delta z$.
Equation \ref{eq:Phibar-hypergeometric-equation} can be solved using the hypergeometric function 
$\hypf (a,b,c;z)$
 \citep[see \textit{e.g.}][]{abramowitz:stegun65,Olver:2010:NHMF} in the form
 \begin{eqnarray}
 \bar{\Phi} =& & \mbox{\hspace{-0.3cm}}A \eta^\delta (1-\eta)^\gamma
\hypf  (\gamma+\delta+1,\gamma+\delta,2\delta+1; \eta) + \nonumber \\
 && \mbox{\hspace{-0.3cm}}B \eta^{-\delta}(1-\eta)^\gamma 
\hypf (\gamma-\delta+1,\gamma-\delta,1-2\delta;\eta),
 \end{eqnarray}
with $\gamma=\sqrt{C_2}$, $\delta=\sqrt{C_1}$,  and $A$, $B$ constants to be determined by boundary conditions at $z=z_{\mathrm{min}}$ (here we take $z_{\mathrm{min}}=0$) and $z= z_{\mathrm{max}}$. 
For example, if we want the solution to tend to zero as $z\to \infty$ ($\eta \to 0$), we have to choose $B=0$. The other coefficient $A$ will be determined by the boundary condition at $z=z_{\mathrm{min}}=0$
in the same way as in the potential or linear force-free magnetic field case. The only difference is the functional dependence of the individual modes on $z$.

We notice that for given values of $a$, $b$ and $\bar{\alpha}$ the parameters $C_1$ and $C_2$ become negative for small $k$ (in particular $C_1 < 0$ if $\bar{k}^2 < \bar{\alpha}^2/[1 -a(1-b)]$ and
$C_2 < 0$ when $\bar{k}^2 < \bar{\alpha}^2/[1 -a(1+b)]$). This implies that $\gamma$, $\delta$ or both become imaginary. The change in nature of the solution is similar to the change from exponential to trigonometric
in the linear force-free case and while this can in principle be incorporated into a Green's function approach \citep[for a discussion see \textit{e.g.}][]{chiu:hilton77, wheatland99, petrie:lothian03, priest14} one usually has to impose
restrictions on the values of $a$, $b$ and $\bar{\alpha}$ in the case with periodic boundary conditions and discrete Fourier modes. As in the linear force-free case the specific bounds are determined by the smallest 
value of the wave vector $k^2$. We will discuss an example in section \ref{sec:examples}.

\section{Illustrative Examples}
\label{sec:examples}

Instead of using an arbitrary magnetogram we
investigate the effects of the solution parameters on the structure of the magnetic field and the plasma pressure, density and temperature, 
by using
a relatively simple, doubly periodic example which 
allows us to control the shape of the bottom boundary conditions and hence enables us to undertake a study of the solutions that highlights the features of the solutions more clearly. 

To achieve this we will use the following boundary condition for $B_z$ at $z=0$:
\begin{eqnarray}
B_z(x,y,0) &=& B_0 \left\{ \frac{\exp\left[ \tilde{\kappa}_{x1} \cos\left(\tilde{x} - \tilde{\mu}_{x1} \right) \right]}{ 2\pi \mbox{I}_0(\tilde{\kappa}_{x1})} 
                                           \frac{\exp\left[ \tilde{\kappa}_{y1} \cos\left(\tilde{y} - \tilde{\mu}_{y1} \right) \right]}{ 2\pi \mbox{I}_0(\tilde{\kappa}_{y1})}  \right.\nonumber \\
                & &    \mbox{\hspace{0.5cm}}    \left.                   -  
                                           \frac{\exp\left[ \tilde{\kappa}_{x2} \cos\left(\tilde{x} - \tilde{\mu}_{x2} \right) \right]}{ 2\pi \mbox{I}_0(\tilde{\kappa}_{x2})} 
                                           \frac{\exp\left[ \tilde{\kappa}_{y2} \cos\left(\tilde{y} - \tilde{\mu}_{y2} \right) \right]}{ 2\pi \mbox{I}_0(\tilde{\kappa}_{y2})}
                                           \right\} .
\label{eq:magnetogram}
\end{eqnarray}
This choice is based on a special case of the bivariate von Mises distribution which is used, 
for example, in directional statistics \citep[see \textit{e.g.}][]{mardia:jupp-book} and we combine two of these functions, one with a positive sign and one with a negative sign, to balance the magnetic flux through the 
lower boundary.
Here $\tilde{x} =\pi x/L$ and $\tilde{y} =\pi y/L$, with the domain size in the $x$- and $y$-directions given by $-1 \le \tilde{x}, \, \tilde{y} \le 1$. All other quantities with a tilde that are directly related
to length scales are also normalised by $L$.
Furthermore, $B_0$ is a reference magnetic field value, the $\tilde{\kappa}_{ij}\,\, (>0) $ values determine the width of the flux distribution, with larger $\tilde{\kappa}_{ij}$ values making the width to smaller, 
the $\tilde{\mu}_{ij}$ values specify the positions of the maximum or minimum of the magnetic flux distribution. The function $\mbox{I}_0(x)$ is a modified Bessel function of the first kind
\citep[see \textit{e.g.}][]{abramowitz:stegun65,Olver:2010:NHMF}. The denominator is included to normalise the integral of the bivariate von Mises distribution over the $x$-$y$ box to unity, so that the positive 
and the negative magnetic flux through the bottom boundary cancel exactly and the total magnetic flux through the bottom boundary vanishes, independently of the values chosen for $\tilde{\kappa}_{ij}$ and
$\tilde{\mu}_{ij}$. The remaining boundary condition imposed is $B_z \to 0$ as $z \to \infty$. 

\begin{figure}
\centerline{\hspace*{0.015\textwidth}
\includegraphics[width=0.515\textwidth,clip=]{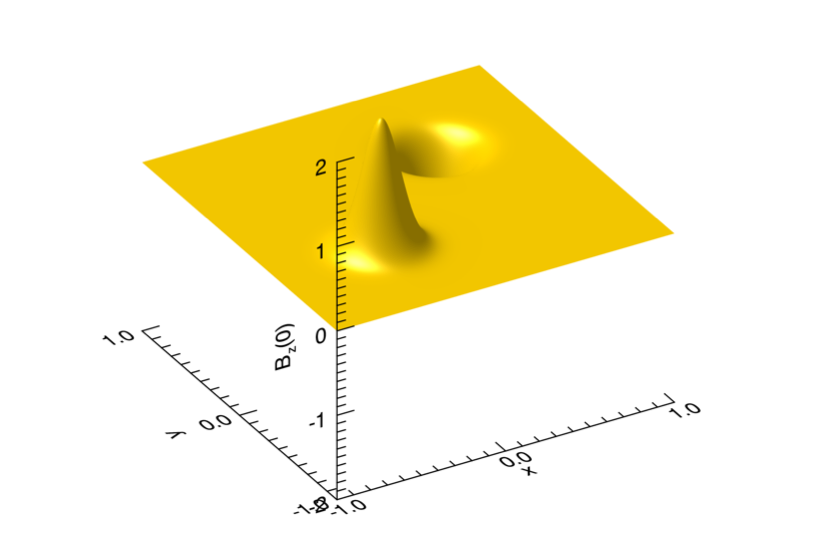}
 \hspace*{-0.03\textwidth}
\includegraphics[width=0.515\textwidth,clip=]{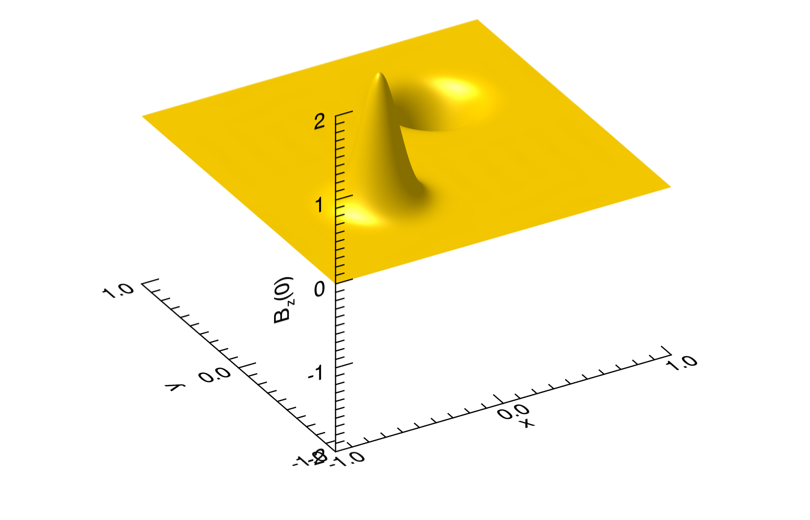}
}
 \vspace{-0.35\textwidth}   
     \centerline{\large \bf     
      \hspace{0.0 \textwidth}  \color{black}{(a)}
      \hspace{0.415\textwidth}  \color{black}{(b)}
         \hfill}
     \vspace{0.31\textwidth}    

\caption{Surface plot of the $B_z$-component of the magnetic field at $z=0$ used for the examples in this paper. Panel (a) shows a plot of the exact expression \ref{eq:magnetogram} and
panel (b) an approximation based on a Fourier expansion using ten Fourier modes. The parameter values used in this plot are 
$\tilde{\mu}_{x1} = \tilde{\mu}_{y1} = - \tilde{\mu}_{x2} = -  \tilde{\mu}_{y2} = 1.2/\pi \approx 0.382 $ and $\kappa_{x1} = \kappa_{x2} = \kappa_{y1} = \kappa_{y2} = 10$.
The maximum value of $|B_z|/B_0$ in this plot is $\approx \exp(2 \tilde{\mu}_{x1}) / [2 \pi \mbox{I}_0(\tilde{\kappa}_{x1})]^2 = 1.55$.
The difference between the exact plot and the ten-mode approximation is of the order of $10^{-5}$.}
\label{fig:magnetogram}
\end{figure}
The function \ref{eq:magnetogram} is periodic in the $x$- and the $y$-direction and can easily be expanded into a Fourier series, which allows us to find the general solution for these boundary conditions
without problems (for mathematical details see Appendix \ref{app:fourier}).
We show surface plots
of the exact expression \ref{eq:magnetogram} and an expansion based on the first ten Fourier modes in both $x$ and $y$ in Figure \ref{fig:magnetogram}. 
The parameter values used in this plot are 
$\tilde{\mu}_{x1} = \tilde{\mu}_{y1} = - \tilde{\mu}_{x2} = -  \tilde{\mu}_{y2} = 1.2/\pi \approx 0.382 $ and $\kappa_{x1} = \kappa_{x2} = \kappa_{y1} = \kappa_{y2} = 10$.
The maximum difference between the exact plot and the ten-mode approximation is of the order of $10^{-5}$ and hence the ten-mode approximation is considered to be sufficiently accurate
for this choice of parameter values.

For the MHS examples we will be showing we have used the values  $z_0 =0.2 L$, $ \Delta z = 0.1 z_0$ and $b=1.0$,
which means that for $z \gg z_0$ the magnetic field tends towards a potential state (in the case $\alpha =0$) or a linear force-free state (if $\alpha \ne 0$). By choosing $b < 1.0$ one could
retain a controlled level of MHS behaviour of the magnetic field above $z_0$.

We will present solutions for different values of the parameter $a$ below the maximum value for $a$,
which is given by
\begin{equation}
a_{\mathrm{max}} = \frac{\bar{k}^2_{\mathrm{min}}-\bar{\alpha}^2}{\bar{k}^2_{\mathrm{min}}(1 + b)}.
\label{eq:amax}
\end{equation}
As discussed in Section \ref{sec:solutions} the condition \ref{eq:amax} follows directly from the definition of $\gamma$ and for given $b$, $\alpha$ and minimum value of $k^2$ corresponds to the
value of $a$ at which $\gamma$ becomes imaginary. One can also see from Equation \ref{eq:amax} that to have $a_{\mathrm{max}} > 0$ one has to have $k^2_{\mathrm{min}} > \alpha^2$ which is the condition for linear force-free field
modes to drop off exponentially with height.

\begin{figure}
\centerline{\hspace*{0.015\textwidth}   
\includegraphics[width=0.515\textwidth,clip=]{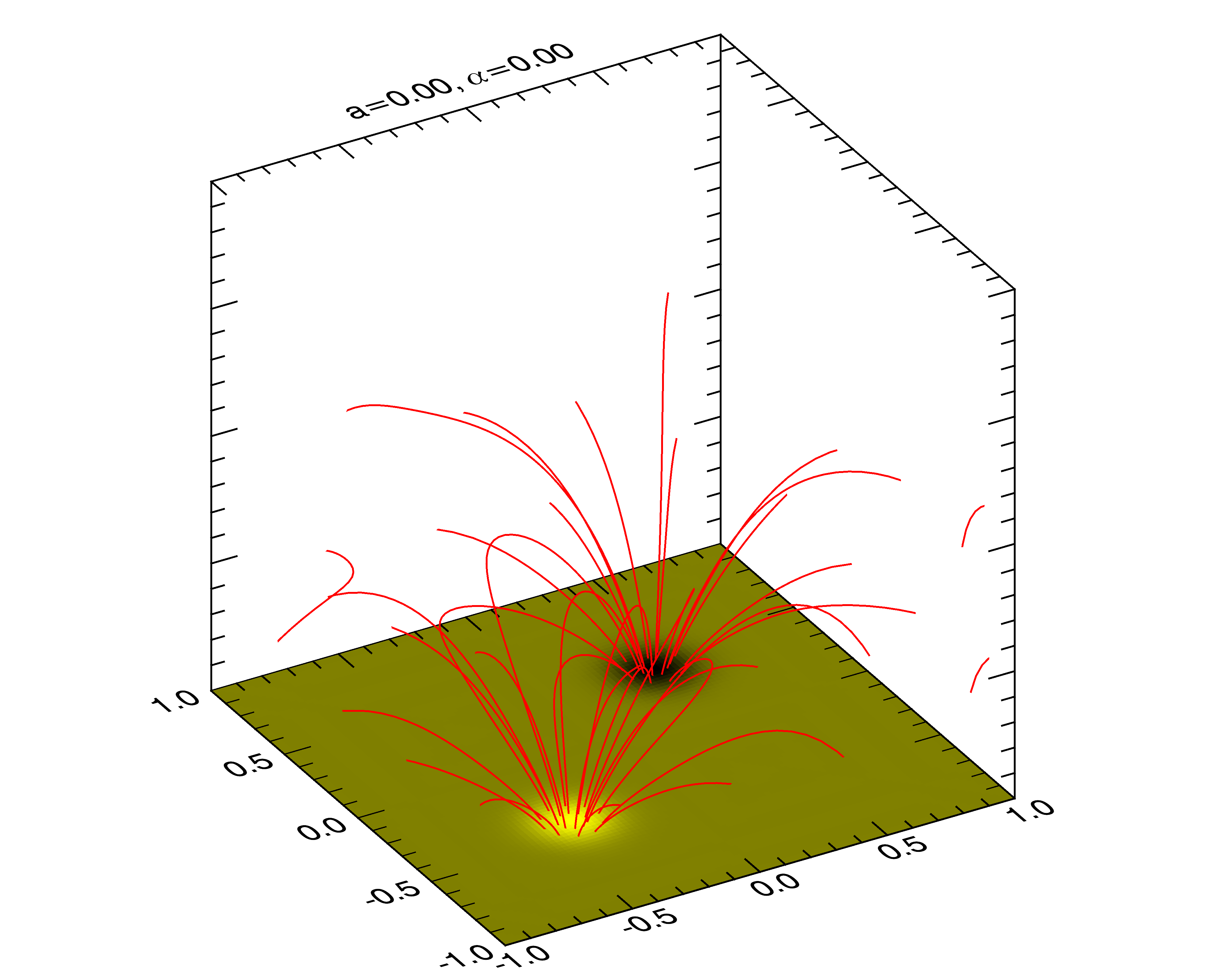}
 \hspace*{-0.03\textwidth}
\includegraphics[width=0.515\textwidth,clip=]{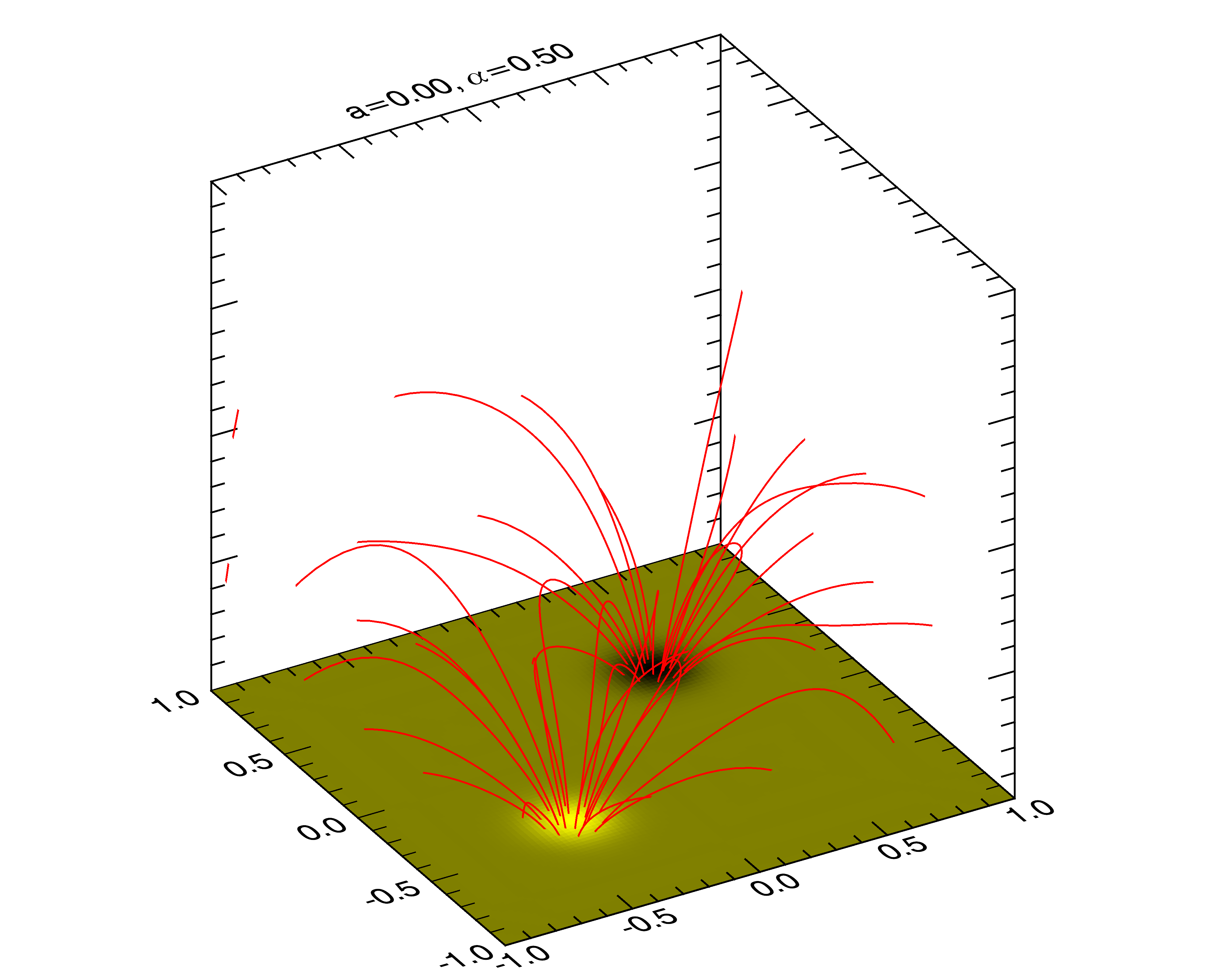}
}
 \vspace{-0.35\textwidth}   
     \centerline{\large \bf     
      \hspace{0.0 \textwidth}  \color{black}{(a)}
      \hspace{0.415\textwidth}  \color{black}{(b)}
         \hfill}
     \vspace{0.31\textwidth}    
\centerline{\hspace*{0.015\textwidth}
\includegraphics[width=0.515\textwidth,clip=]{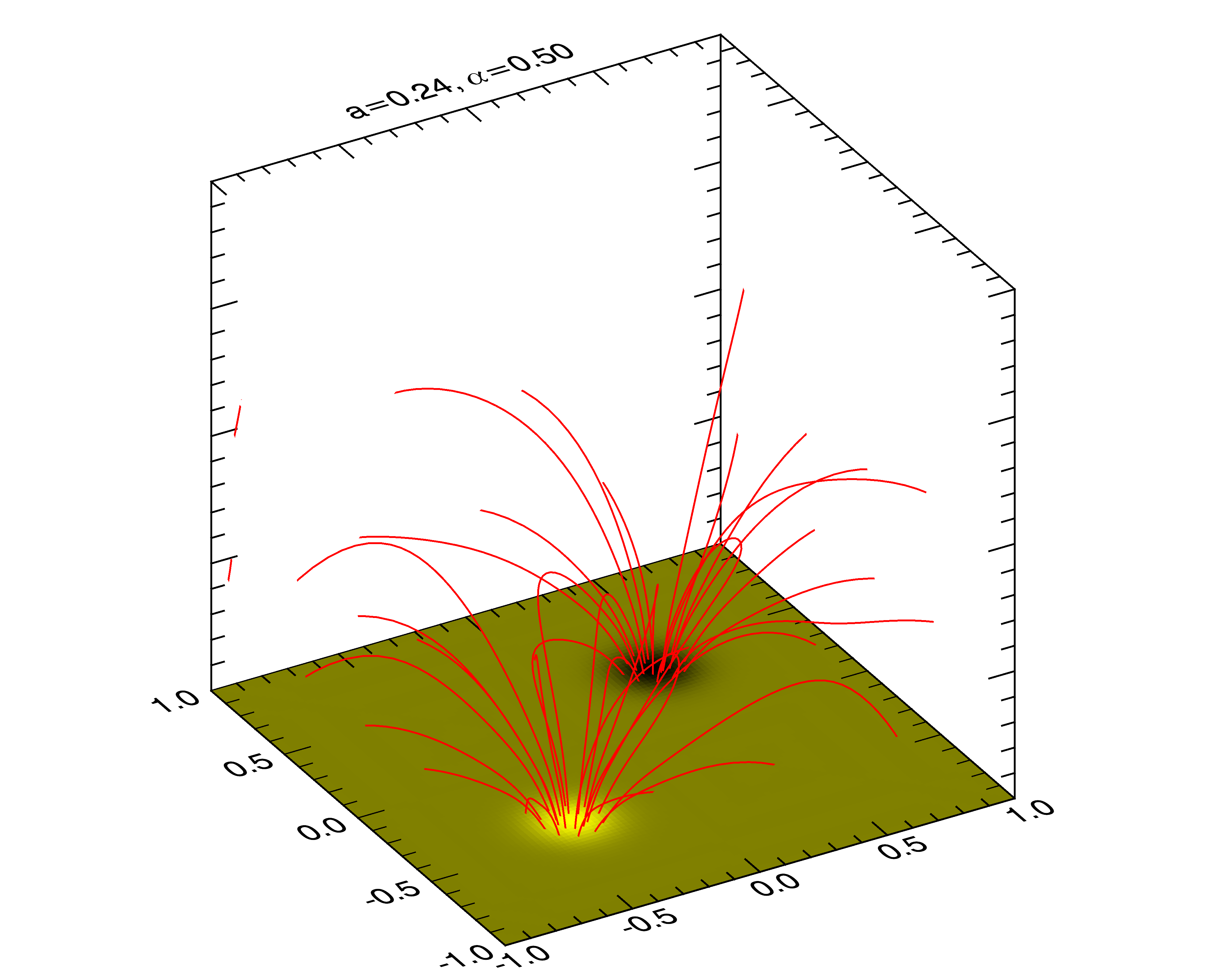}
 \hspace*{-0.03\textwidth}
\includegraphics[width=0.515\textwidth,clip=]{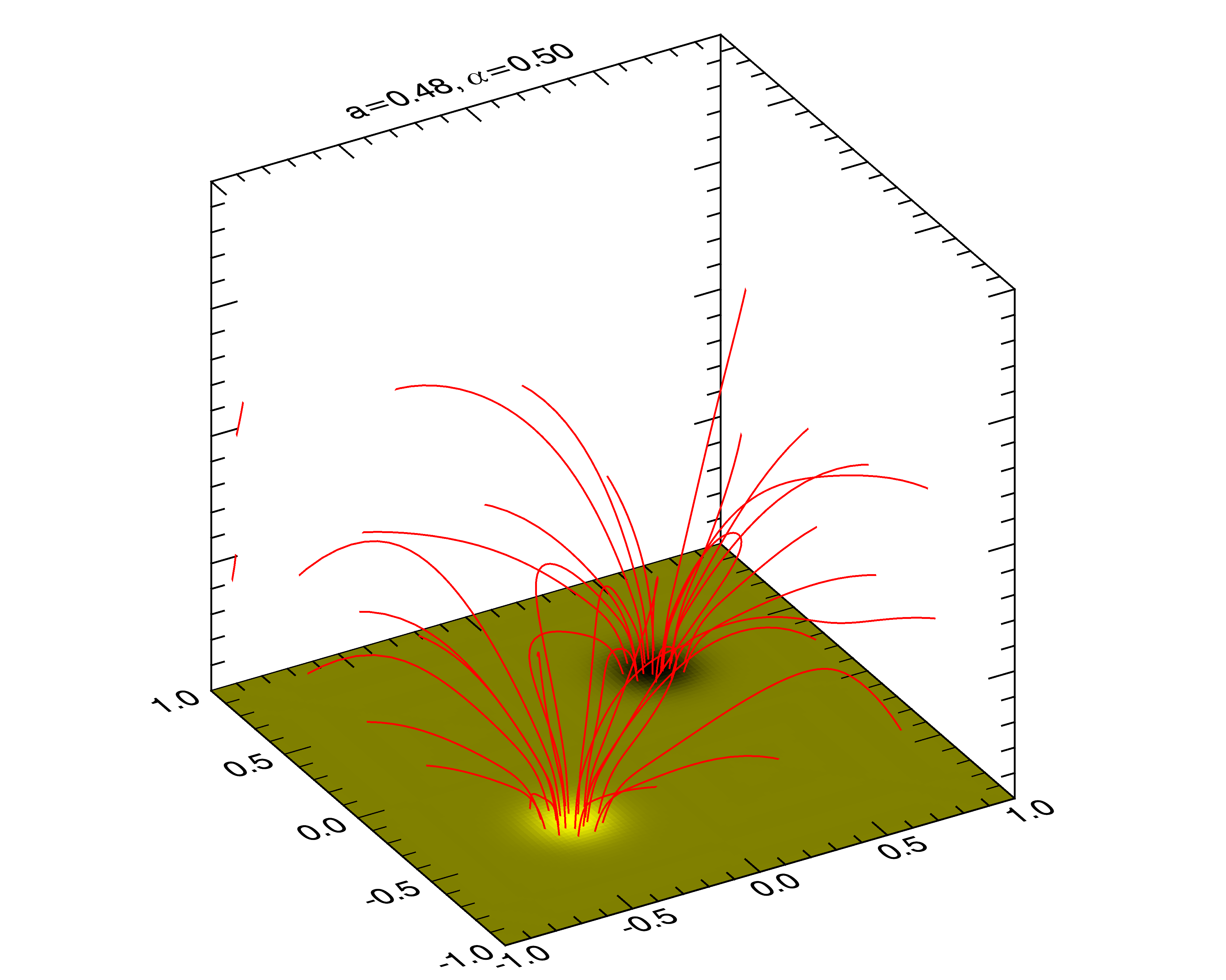}
}
 \vspace{-0.35\textwidth}   
     \centerline{\large \bf     
      \hspace{0.0 \textwidth}  \color{black}{(c)}
      \hspace{0.415\textwidth}  \color{black}{(d)}
         \hfill}
     \vspace{0.31\textwidth}    
     
 \caption{Field line plots for the boundary conditions shown in Figure \ref{fig:magnetogram}. The box is
 a cube with size is $-1 \le x/L,\, y/L \le 1$,
 $0 \le z/L \le 2$.
 Panel (a) shows the potential magnetic field ($a=0$, $\tilde{\alpha}=0$), panel (b) a linear force-free magnetic field 
 ($a=0$, $\tilde{\alpha}=0.5$) and panels (c) and (d) two MHS solutions with $a=0.24$, $\tilde{\alpha}=0.5$, in panel (c)
 and $a=0.48$, $\tilde{\alpha}=0.5$, in panel (d). In each panel we show 
field lines traced from the same foot points located on two concentric circles around the position of the positive polarity maximum.
The field lines are traced until the encounter the lower boundary, with the periodic boundary conditions are taken into account.}
   \label{fig:B-perspective}
\end{figure}
\begin{figure}                    
\centerline{\hspace*{0.015\textwidth}   
\includegraphics[width=0.45\textwidth, bb = 110 60 360 320, clip=]{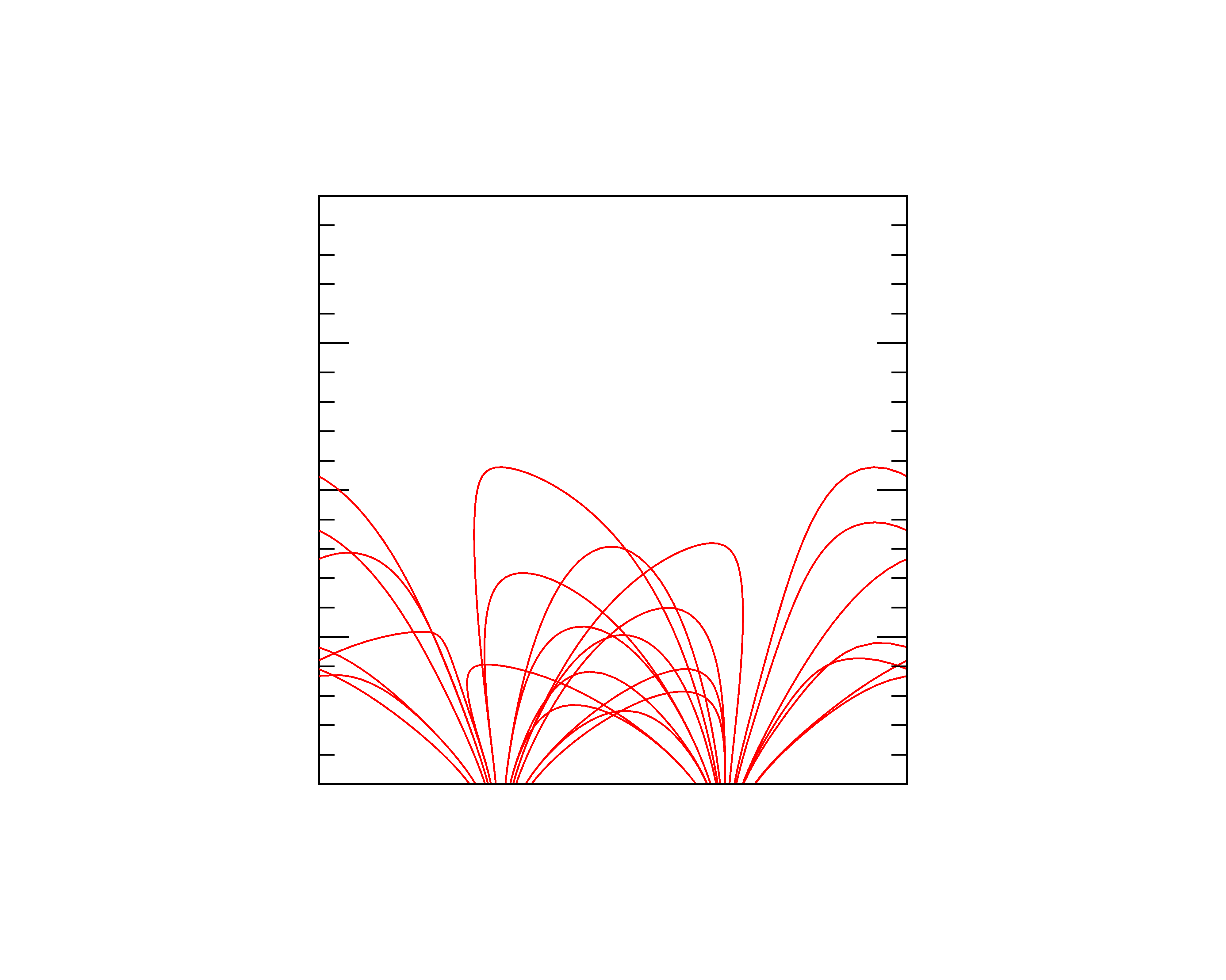}
 \hspace*{0.01\textwidth}
\includegraphics[width=0.45\textwidth,  bb = 110 60 360 320, clip=]{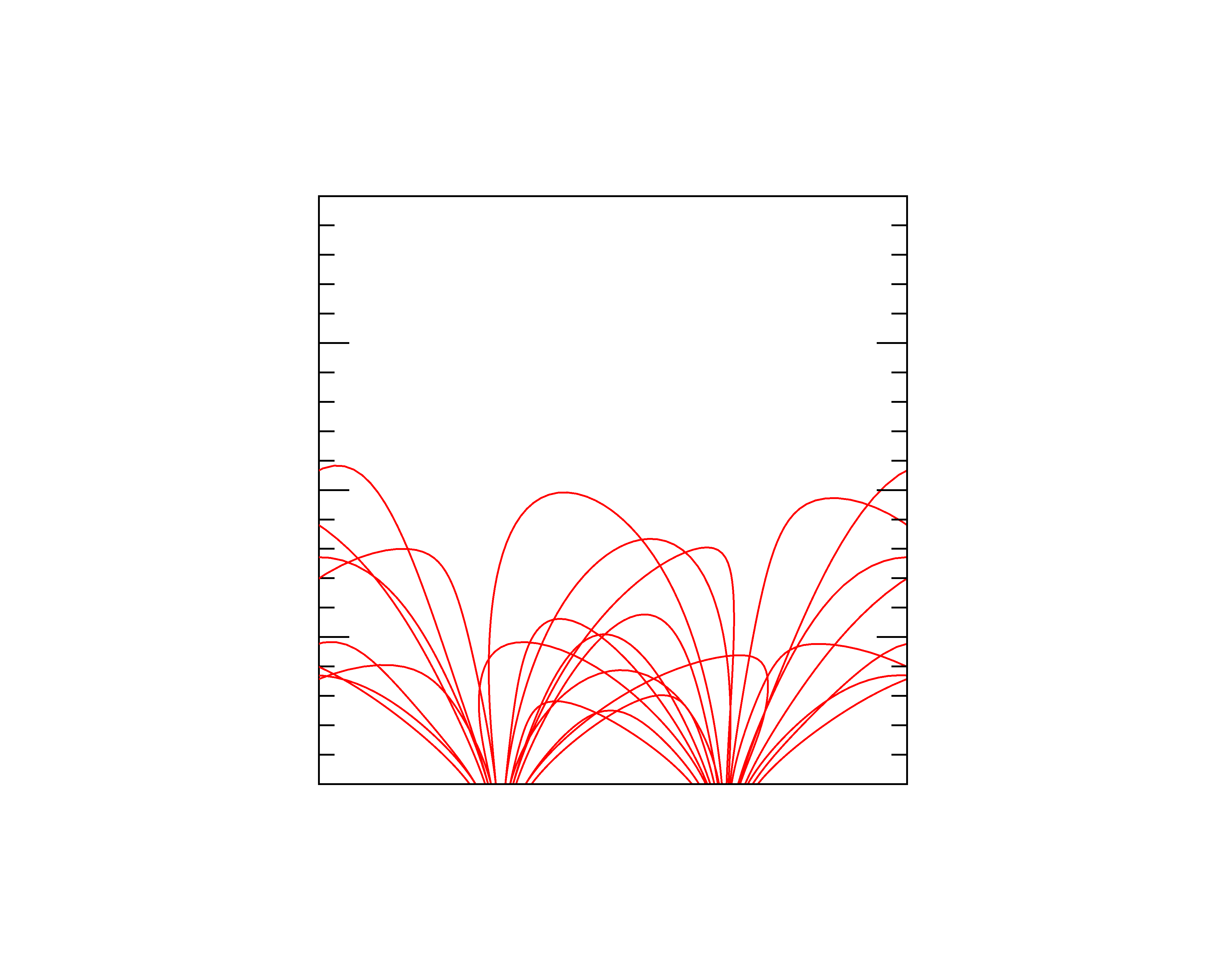}
}
 \vspace{-0.42\textwidth}   
     \centerline{\large \bf     
      \hspace{0.075 \textwidth}  \color{black}{(a)}
      \hspace{0.41\textwidth}  \color{black}{(b)}
         \hfill}
     \vspace{0.42\textwidth}    
\centerline{\hspace*{0.015\textwidth}
\includegraphics[width=0.45\textwidth, bb = 110 60 360 320, clip=]{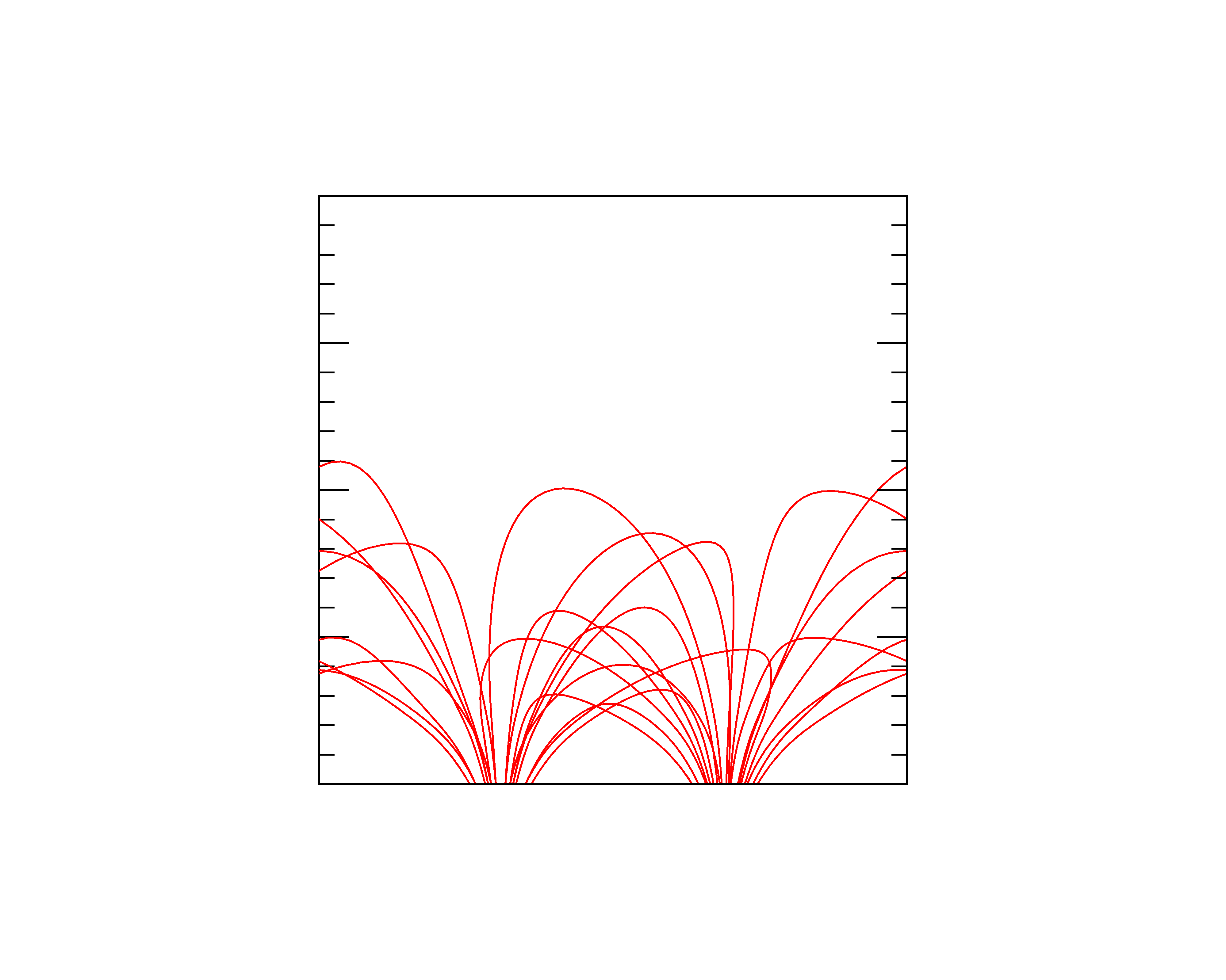}
 \hspace*{0.01\textwidth}
\includegraphics[width=0.45\textwidth, bb = 110 60 360 320, clip=]{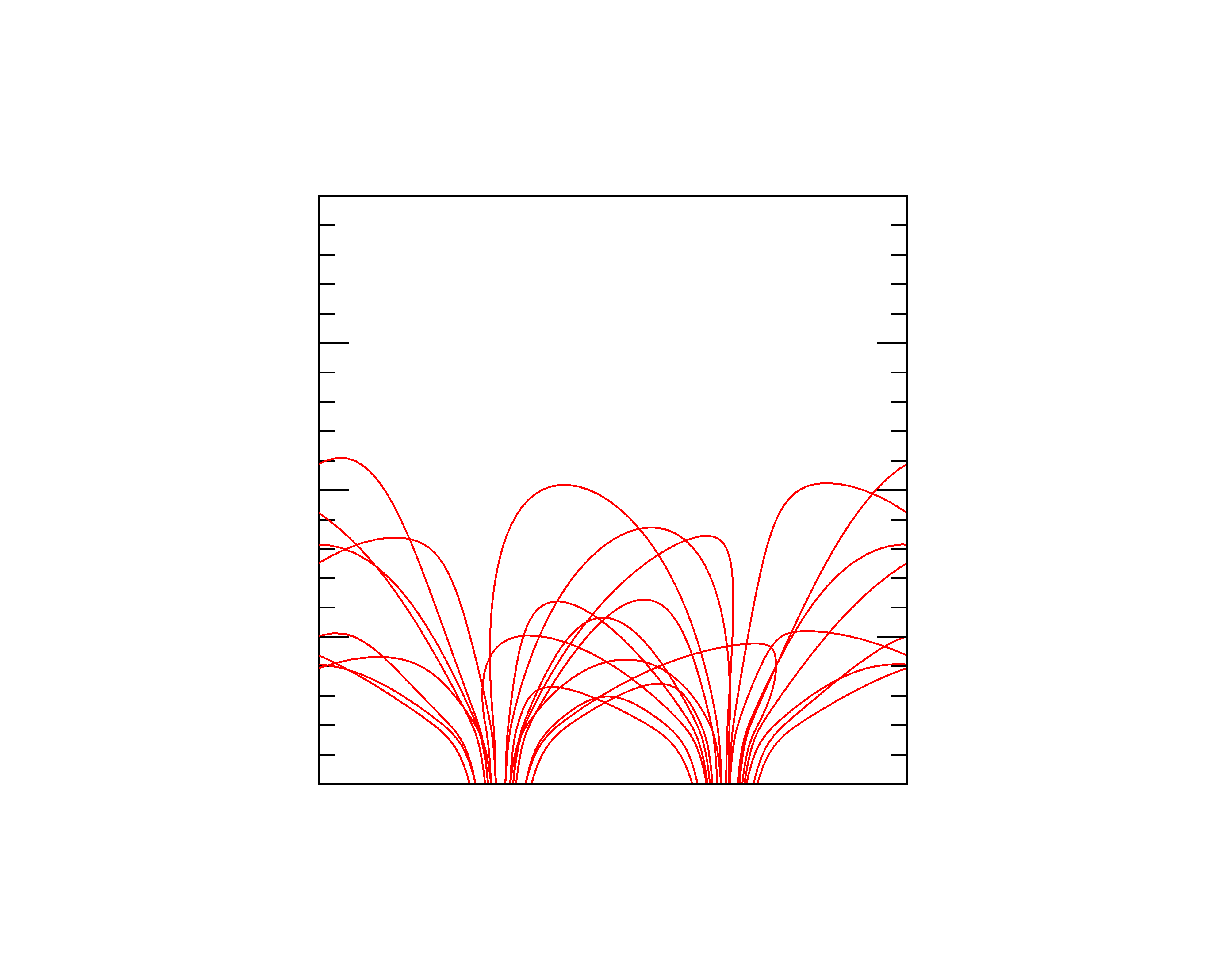}
}
 \vspace{-0.42\textwidth}   
     \centerline{\large \bf     
      \hspace{0.075 \textwidth}  \color{black}{(c)}
      \hspace{0.41\textwidth}  \color{black}{(d)}
         \hfill}
     \vspace{0.42\textwidth}    
     
 \caption{The same field line plots as in Figure \ref{fig:B-perspective}, but viewed along the $y$-direction ({\it i.e.}\ projected onto the $x$-$z$-plane).
 As in Figure \ref{fig:B-perspective} we have $-1 \le x/L \le 1$ (horizontal direction) and $0 \le z/L \le 2$ (vertical direction).
The field lines in the MHS cases (panels (c) and (d)) show clear signs of steepening below $z/L=z_0/L=0.2$ (second tickmark on the vertical $z$-axis), as expected.}
   \label{fig:B-sideview}
\end{figure}

\begin{figure}
\centerline{\hspace*{0.0\textwidth}   
\includegraphics[width=0.6\textwidth,clip=]{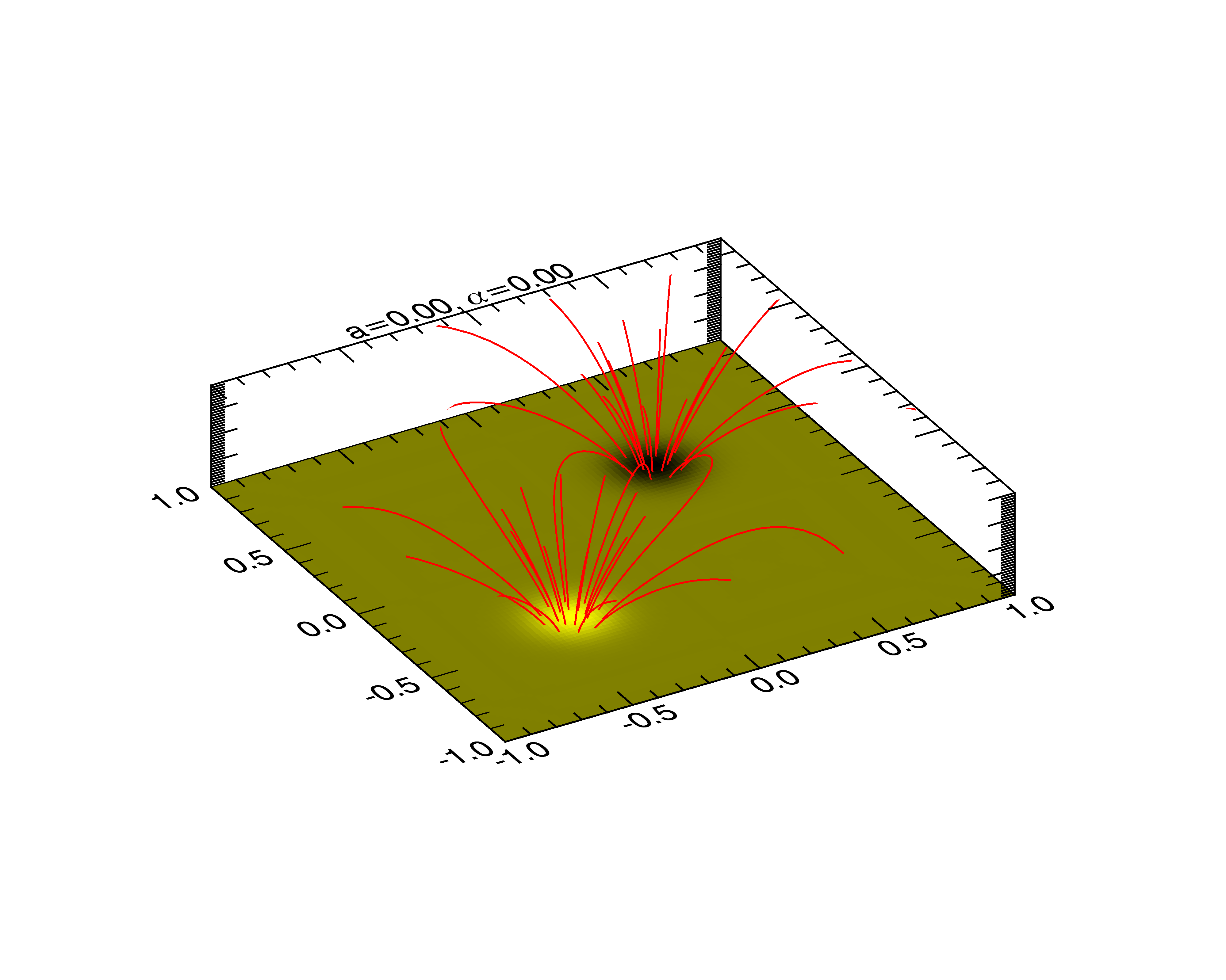}
 \hspace*{-0.13\textwidth}
\includegraphics[width=0.6\textwidth,clip=]{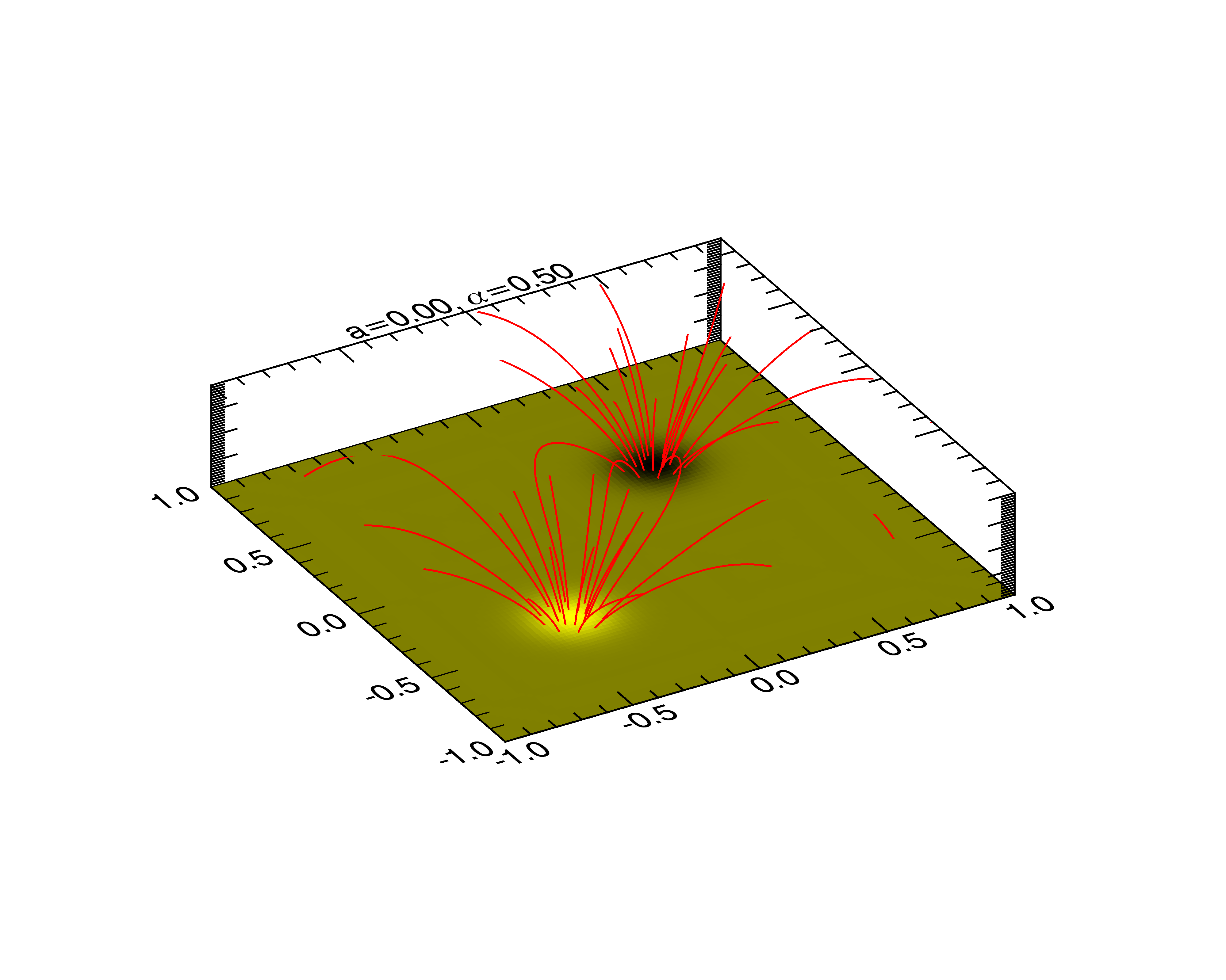}
}
 \vspace{-0.42\textwidth}   
     \centerline{\large \bf     
      \hspace{0.0 \textwidth}  \color{black}{(a)}
      \hspace{0.5\textwidth}  \color{black}{(b)}
         \hfill}
     \vspace{0.3\textwidth}    
\centerline{\hspace*{0.05\textwidth}
\includegraphics[width=0.6\textwidth,clip=]{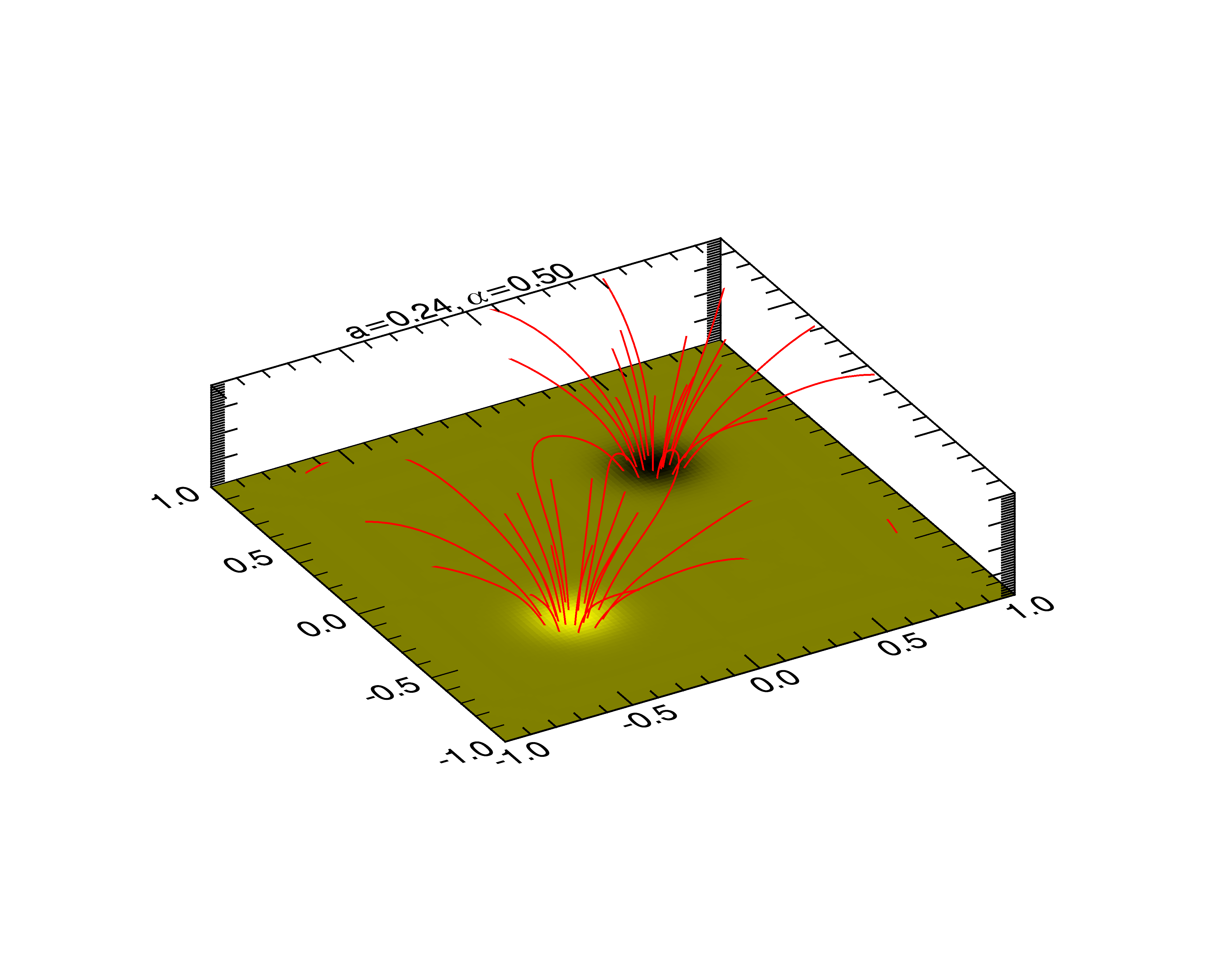}
 \hspace*{-0.13\textwidth}
\includegraphics[width=0.6\textwidth,clip=]{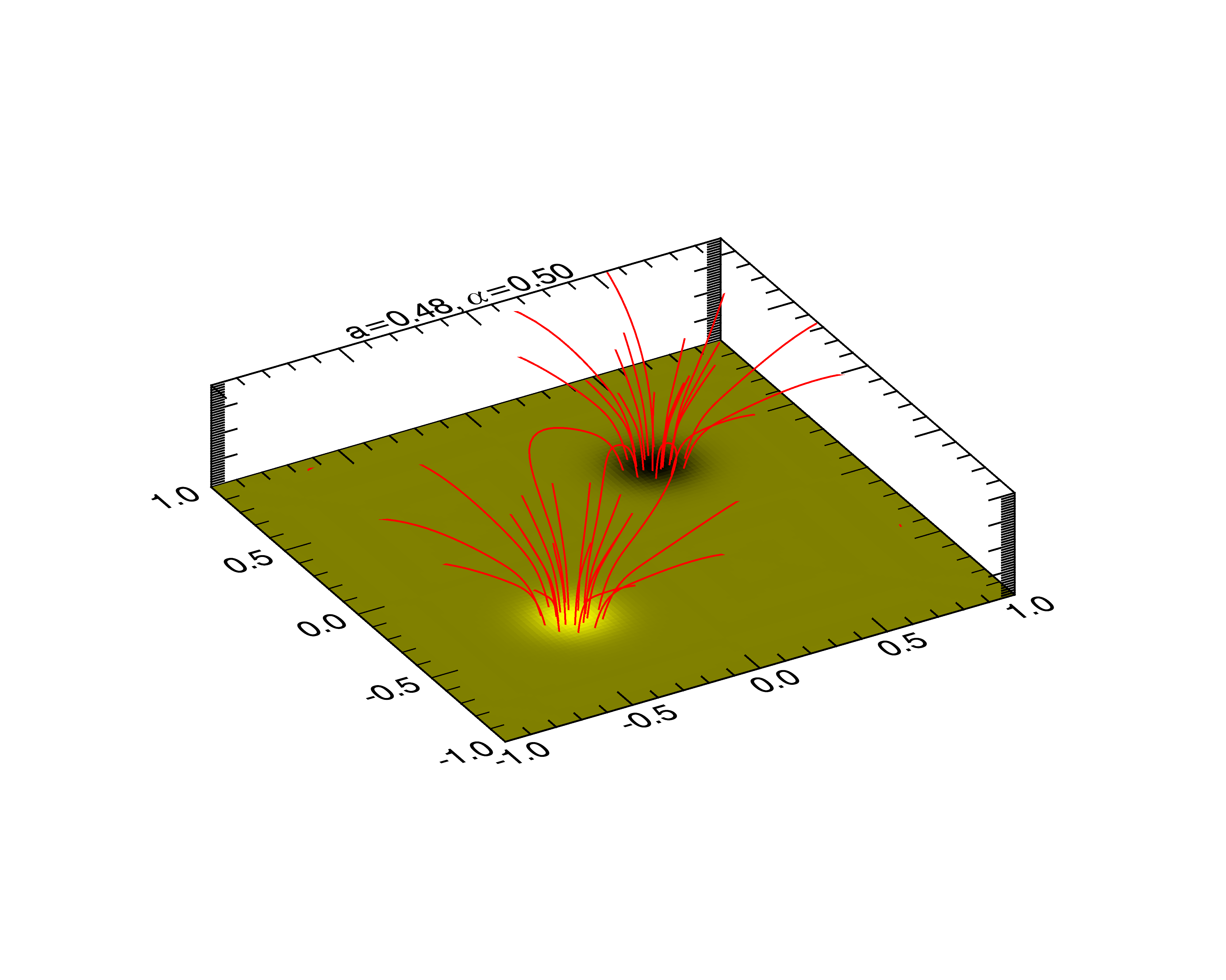}
}
 \vspace{-0.42\textwidth}   
     \centerline{\large \bf     
      \hspace{0.0 \textwidth}  \color{black}{(c)}
      \hspace{0.5\textwidth}  \color{black}{(d)}
         \hfill}
     \vspace{0.37\textwidth}    
     
 \caption{The same field line plots as in Figure \ref{fig:B-perspective}, but with $0 \le z \le 2 z_0$ to make the difference between the various cases more obvious.}
   \label{fig:B-perspective-lowb}
\end{figure}
\begin{figure}                    
\centerline{\hspace*{0.0\textwidth}   
\vspace{0.1cm}
\includegraphics[width=0.45\textwidth, bb = 110 120 360 240, clip=]{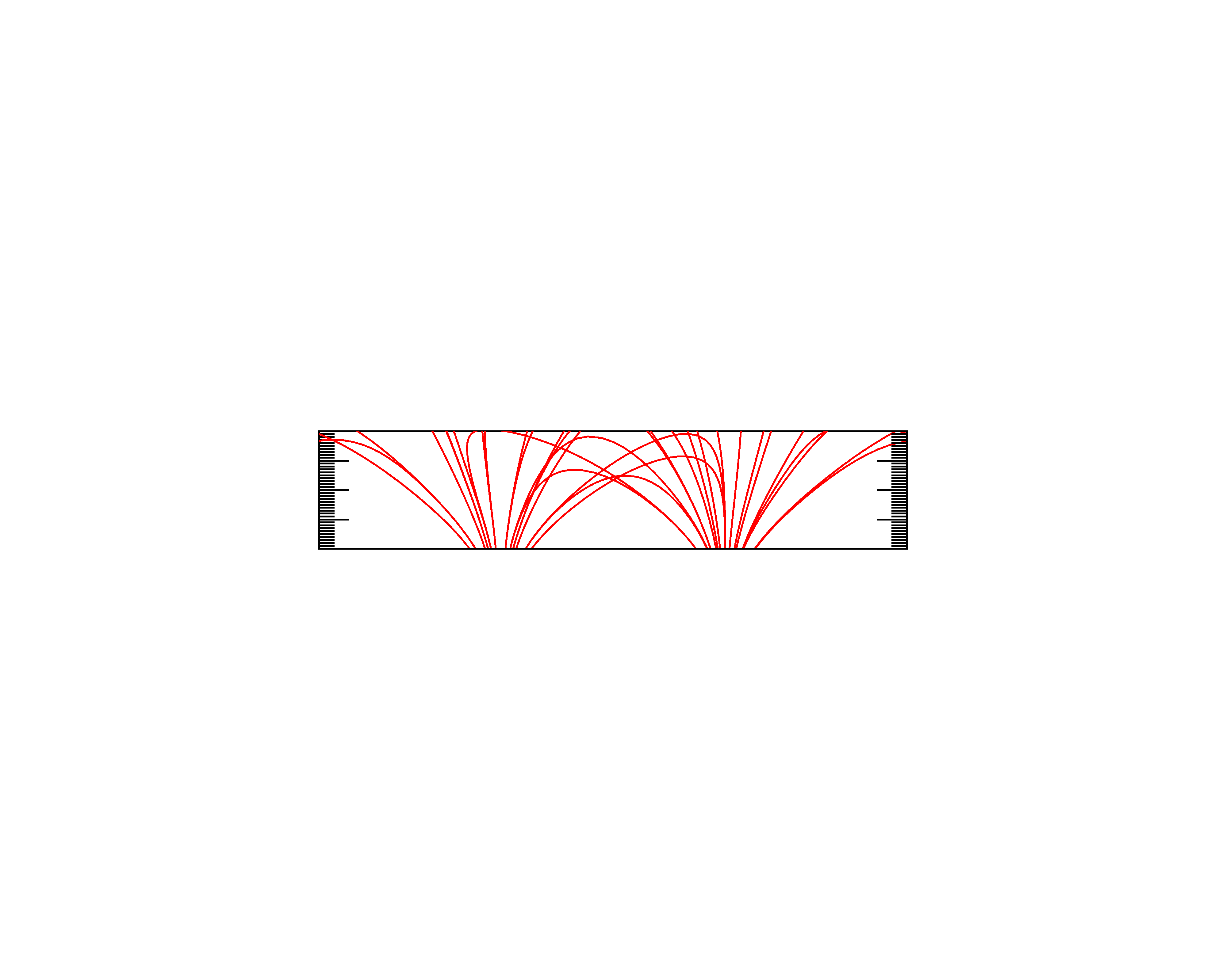}
 \hspace*{-0.0\textwidth}
\includegraphics[width=0.45\textwidth,  bb = 110 120 360 240, clip=]{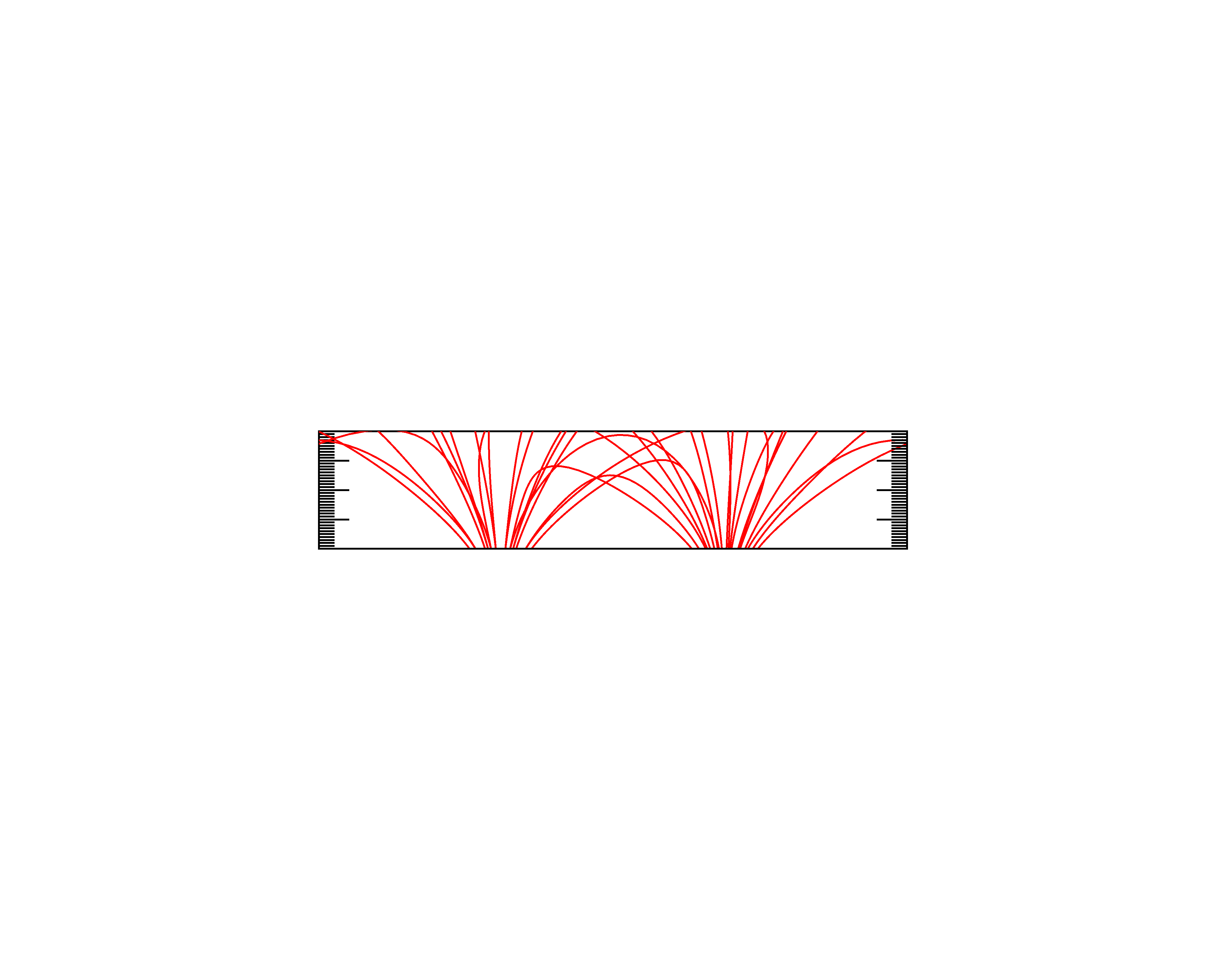}
}
 \vspace{-0.24\textwidth}   
     \centerline{\large \bf     
      \hspace{0.05 \textwidth}  \color{black}{(a)}
      \hspace{0.4\textwidth}  \color{black}{(b)}
         \hfill}
     \vspace{0.15\textwidth}    
\centerline{\hspace*{0.0\textwidth}
\includegraphics[width=0.45\textwidth, bb = 110 120 360 240, clip=]{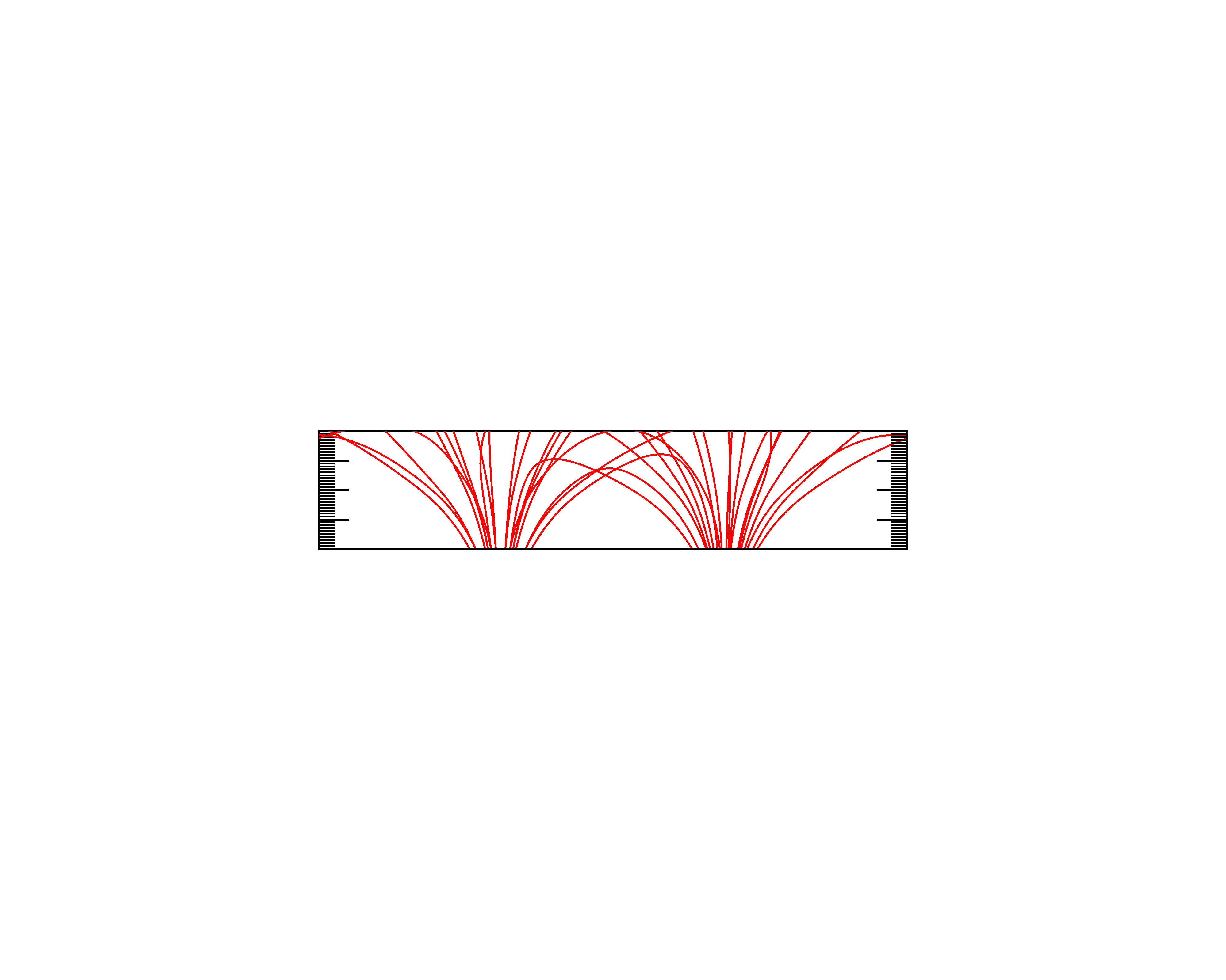}
 \hspace*{-0.0\textwidth}
\includegraphics[width=0.45\textwidth, bb = 110 120 360 240, clip=]{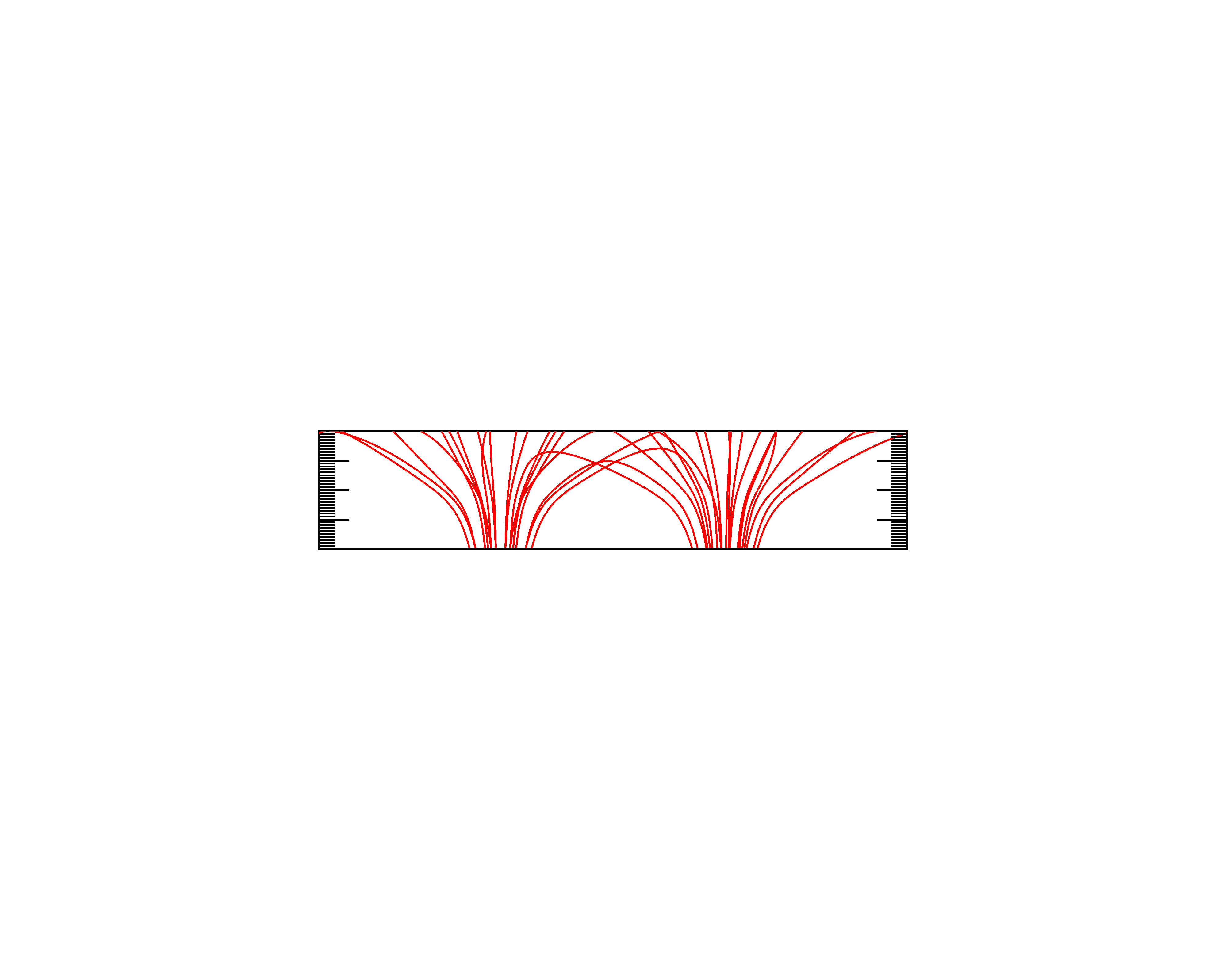}
}

 \vspace{-0.23\textwidth}   
     \centerline{\large \bf     
      \hspace{0.05 \textwidth}  \color{black}{(c)}
      \hspace{0.4\textwidth}  \color{black}{(d)}
         \hfill}
     \vspace{0.15\textwidth}    
     
 \caption{The same field line plots as in Figure \ref{fig:B-perspective-lowb}, but viewed along the $y$-direction ({\it i.e.}\ projected onto the $x$-$z$-plane).}
   \label{fig:B-sideview-lowb}
\end{figure}

We present magnetic field line plots for four different parameter combinations in Figures \ref{fig:B-perspective} and \ref{fig:B-sideview}. For reference we present the potential magnetic field for the
given boundary conditions ($a=0.0$, $\alpha=0.0$) in panel (a) and the linear force-free magnetic field with $\alpha=0.5$ ($a=0$) in panel (b)
of Figures \ref{fig:B-perspective} and \ref{fig:B-sideview}.
We compare these two cases with two field line plots for non-zero values of $a$, one roughly half of $a_{\mathrm{max}}$ and the other $0.99 \, a_{\mathrm{max}}$. 
To highlight the region where the field lines change the most between the different cases, we also present the same field line plots on a domain with $z \le 2 z_0$ in Figures\ \ref{fig:B-perspective-lowb} and
\ref{fig:B-sideview-lowb}.

As expected, the main difference between the linear force-free case and the two MHS cases can be seen for $z \le z_0$. This is particularly obvious in Figure \ref{fig:B-sideview}, in which
a considerable steepening of the field lines is noticeable in the region below $z_0=0.2$. This change in the field line behaviour can be attributed to the value of the smallest $\gamma$ in the series expansion
approaching zero which leads to a less rapid decrease of the lowest order mode with height for $z \le z_0$. Due to our choice of a relatively small value for $\Delta z$ this behaviour changes relatively
sharply around $z \approx z_0$ and hence no major changes in the magnetic field can be seen at heights above $z_0$. The magnetic field above $z_0$ is therefore basically identical to a linear force-free 
field.

\begin{figure}
\centerline{\hspace*{0.025\textwidth}
\includegraphics[width=0.7\textwidth,clip=]{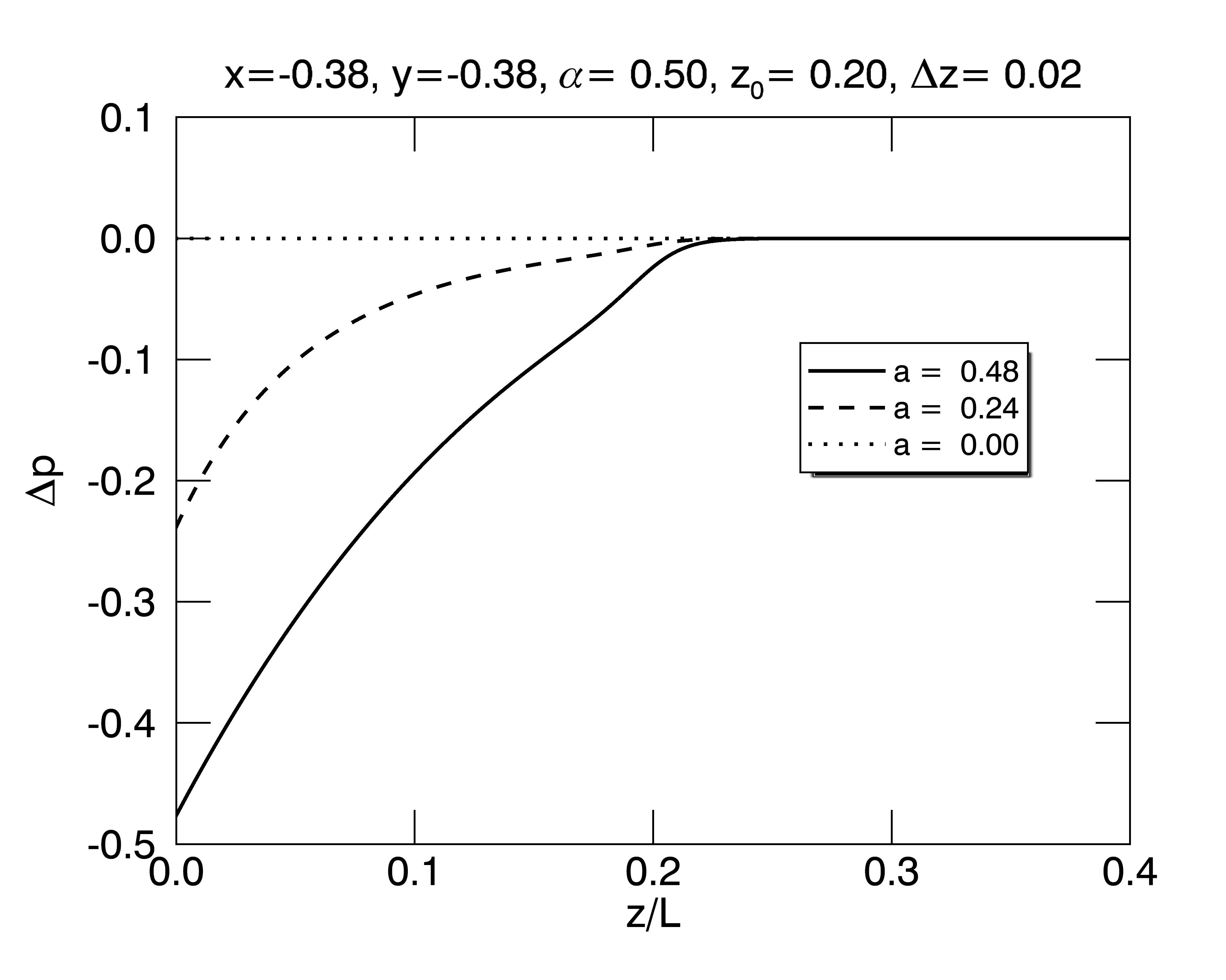}
}
 \vspace{-0.5\textwidth}   
     \centerline{\large \bf     
      \hspace{0.1\textwidth}  \color{black}{(a)}
         \hfill}
     \vspace{0.5\textwidth}    
 \hspace*{-0.03\textwidth}
\centerline{\hspace*{0.025\textwidth}
\includegraphics[width=0.7\textwidth,clip=]{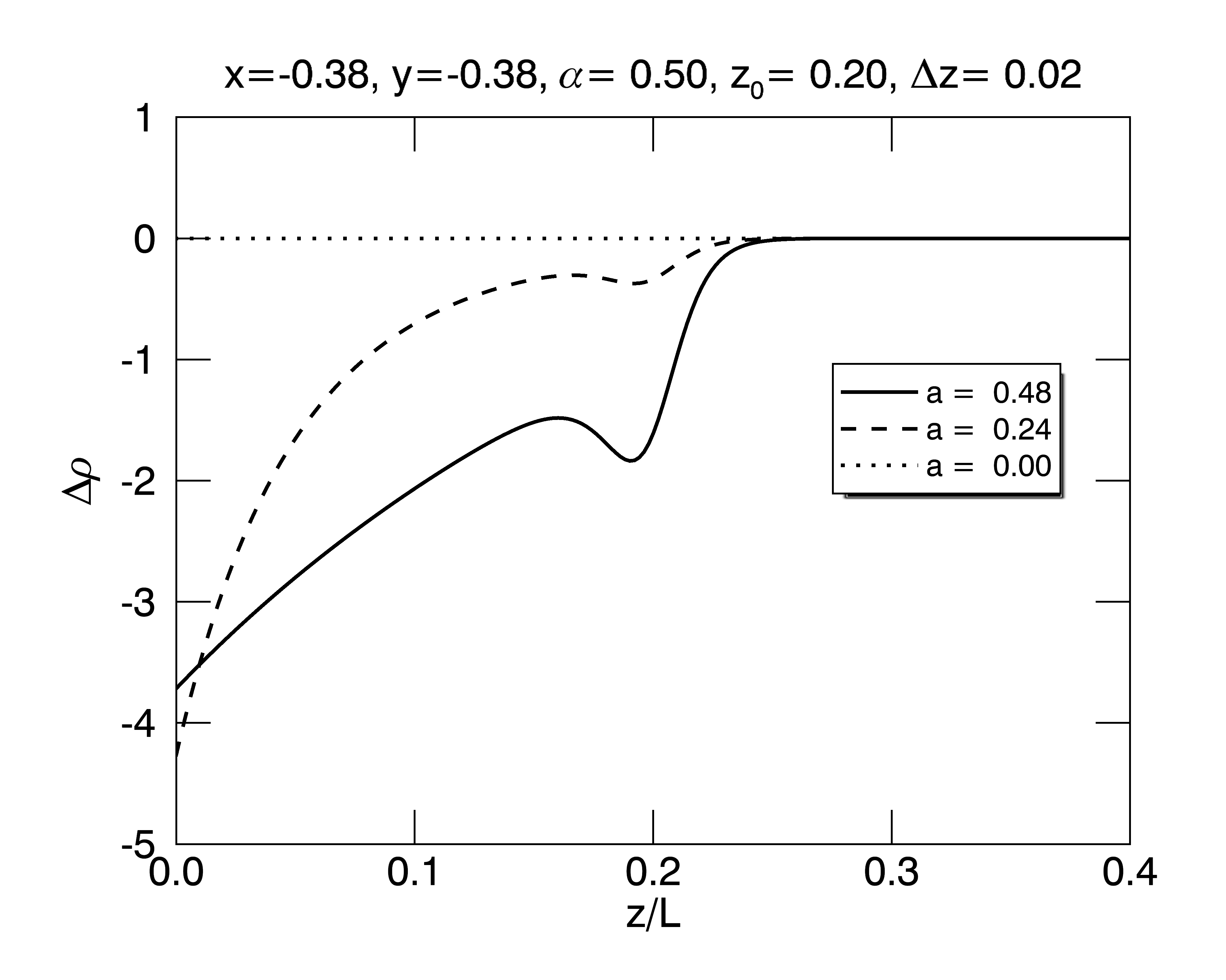}
}

 \vspace{-0.5\textwidth}   
     \centerline{\large \bf     
      \hspace{0.1\textwidth}  \color{black}{(b)}
         \hfill}
     \vspace{0.5\textwidth}    
   \caption{Plots of the variation with height $z$ of the pressure (panel (a)) and density (panel(b)) deviation from a stratified background atmosphere 
   for $\alpha = 0.5$, $z_0/L = 0.2$ and $\Delta z = 0.1 z_0$, and
   three different values of $a$ ($a = 0.0$, $0.24$ and $0.48$). The $x$- and $y$-coordinates of these plots are the position of the maximum value of $|B_z|$ on the lower boundary, 
   $x/L=\mu_x/\pi \approx -0.38$, $y/L=\mu_y/\pi \approx -0.38$.}
   \label{fig:pandrho-lineplots}
\end{figure}

\begin{figure}
\centerline{\hspace*{0.015\textwidth}   
\includegraphics[width=0.515\textwidth,clip=]{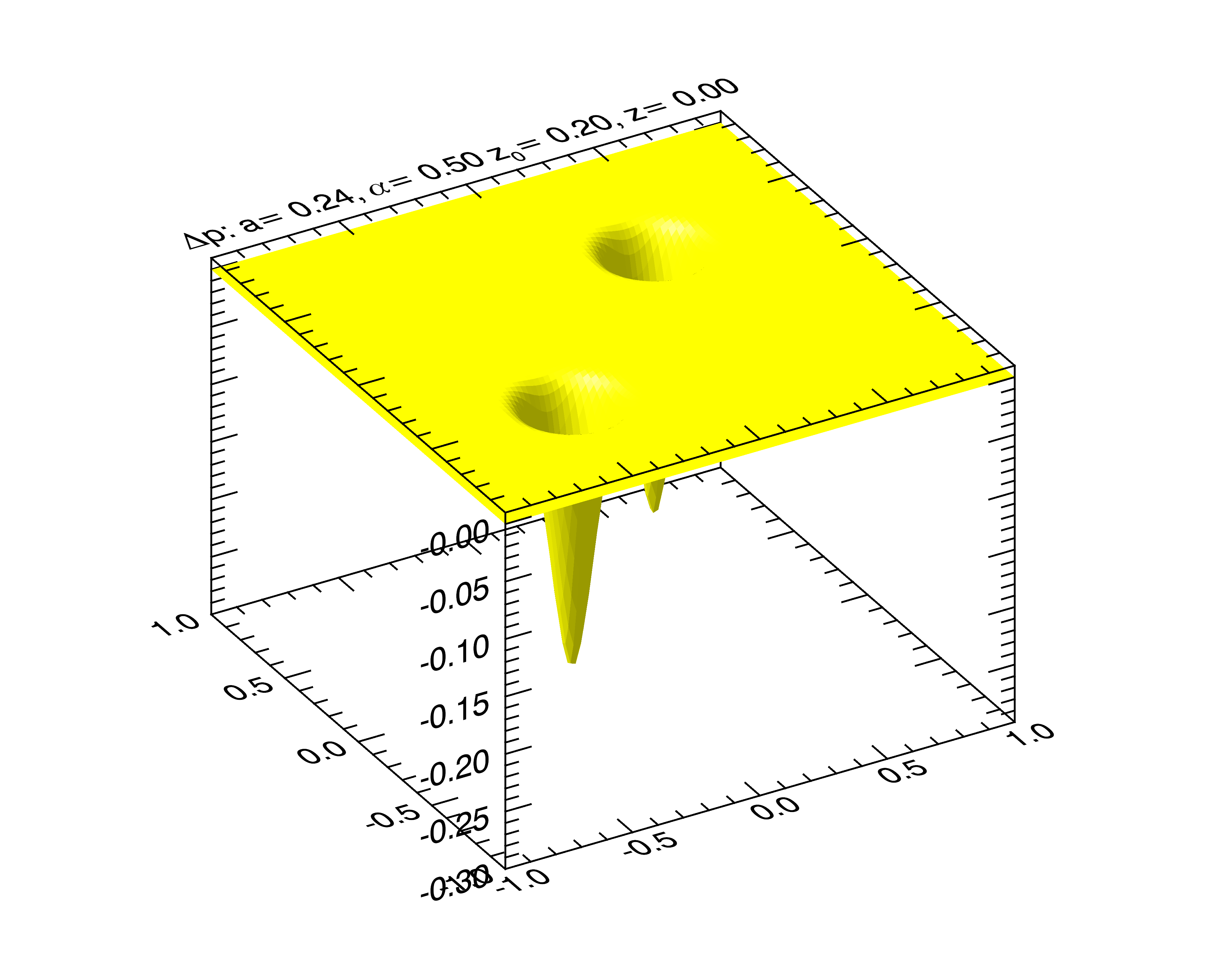}
 \hspace*{-0.03\textwidth}
\includegraphics[width=0.515\textwidth,clip=]{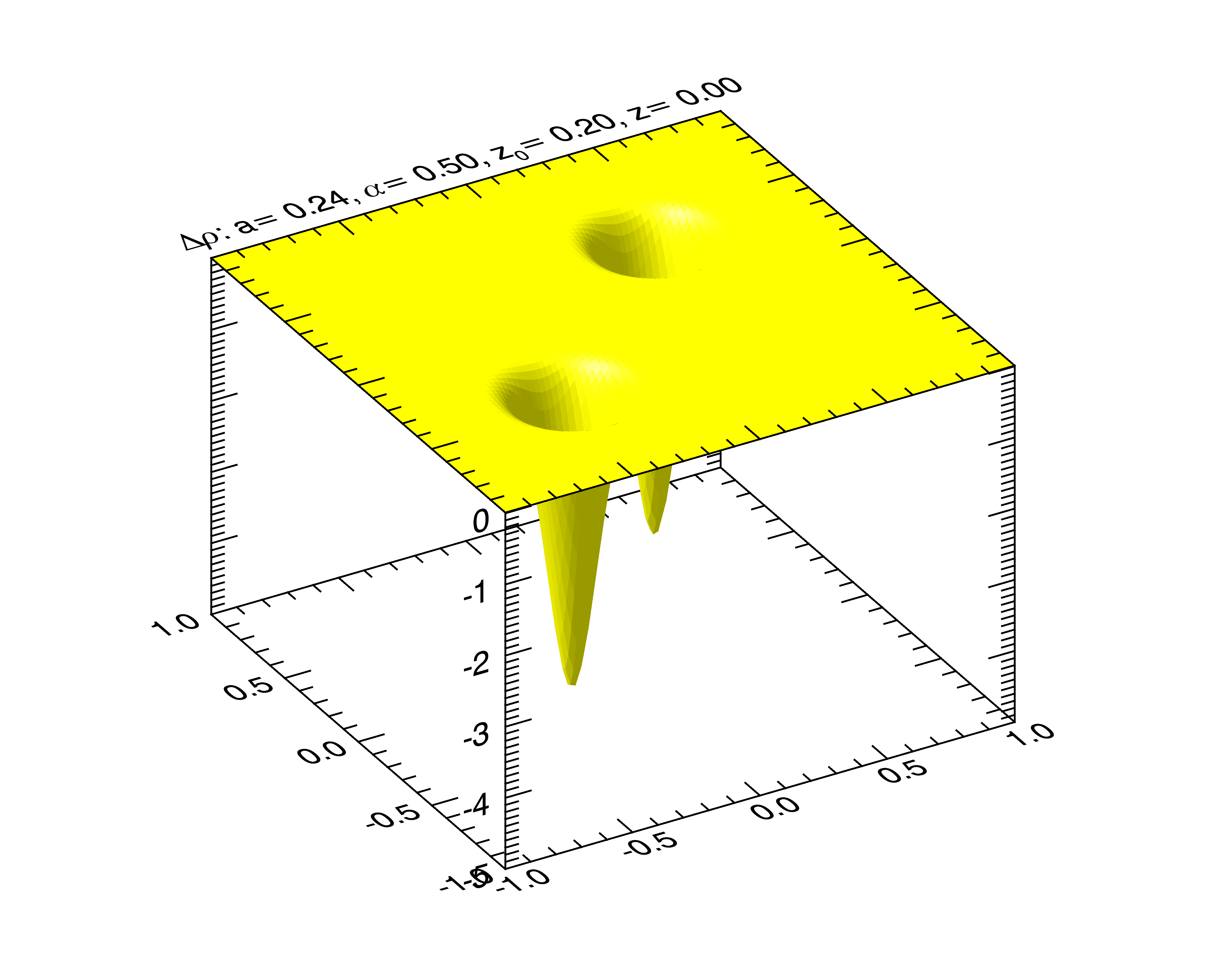}
}
 \vspace{-0.35\textwidth}   
     \centerline{\large \bf     
      \hspace{0.0 \textwidth}  \color{black}{(a)}
      \hspace{0.415\textwidth}  \color{black}{(b)}
         \hfill}
     \vspace{0.31\textwidth}    
\centerline{\hspace*{0.015\textwidth}
\includegraphics[width=0.515\textwidth,clip=]{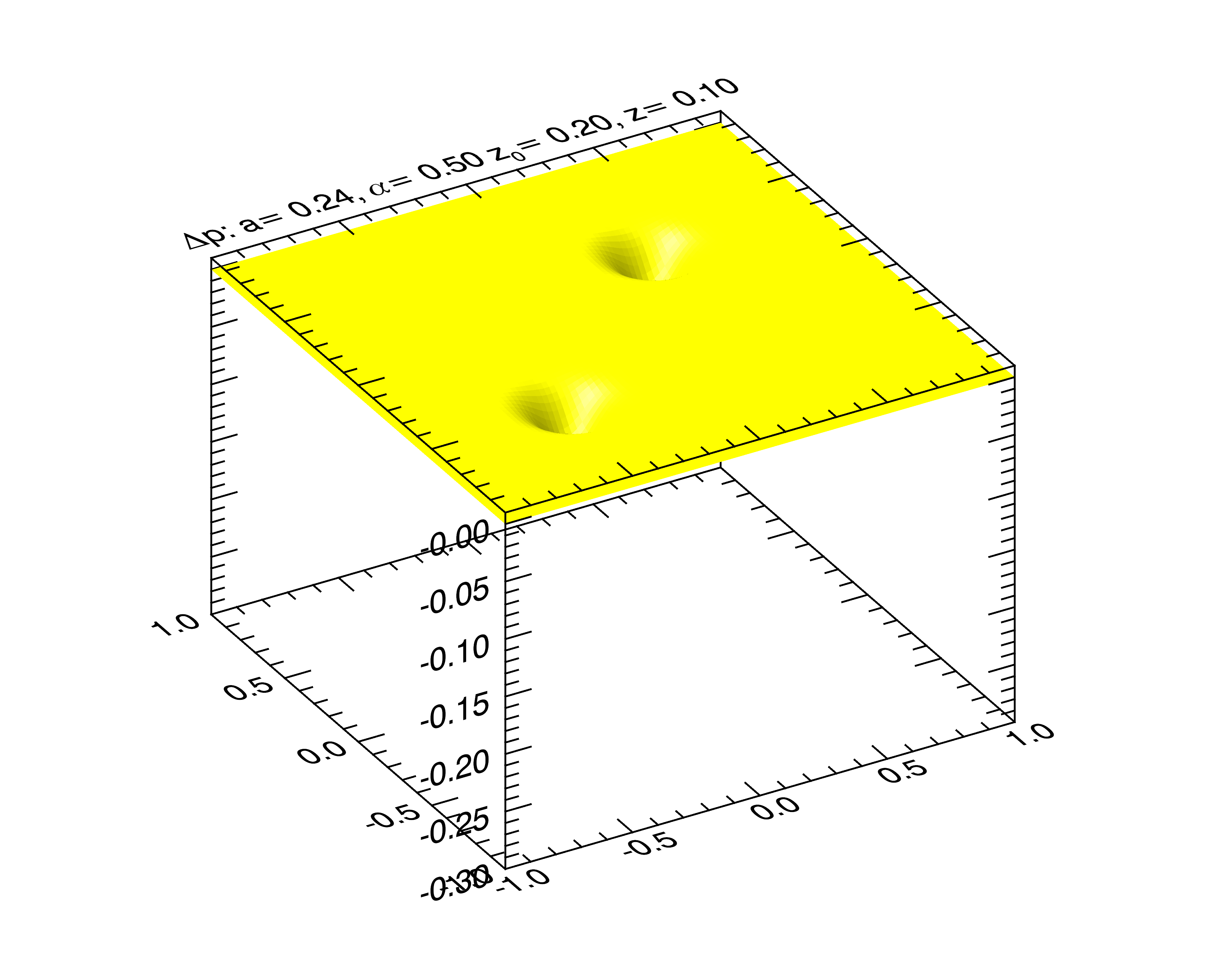}
 \hspace*{-0.03\textwidth}
\includegraphics[width=0.515\textwidth,clip=]{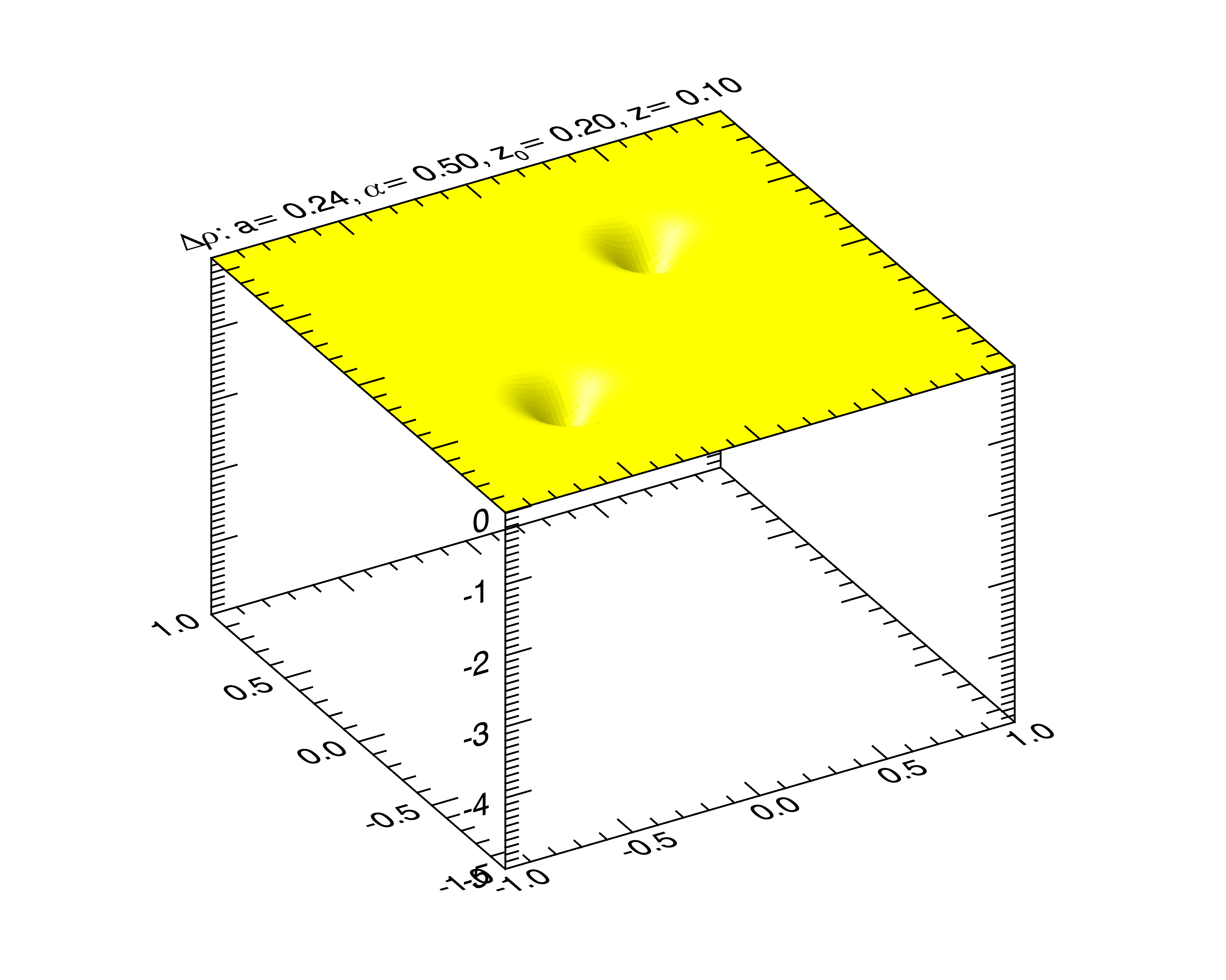}
}
 \vspace{-0.35\textwidth}   
     \centerline{\large \bf     
      \hspace{0.0 \textwidth}  \color{black}{(c)}
      \hspace{0.415\textwidth}  \color{black}{(d)}
         \hfill}
     \vspace{0.31\textwidth}    
\centerline{\hspace*{0.015\textwidth}
\includegraphics[width=0.515\textwidth,clip=]{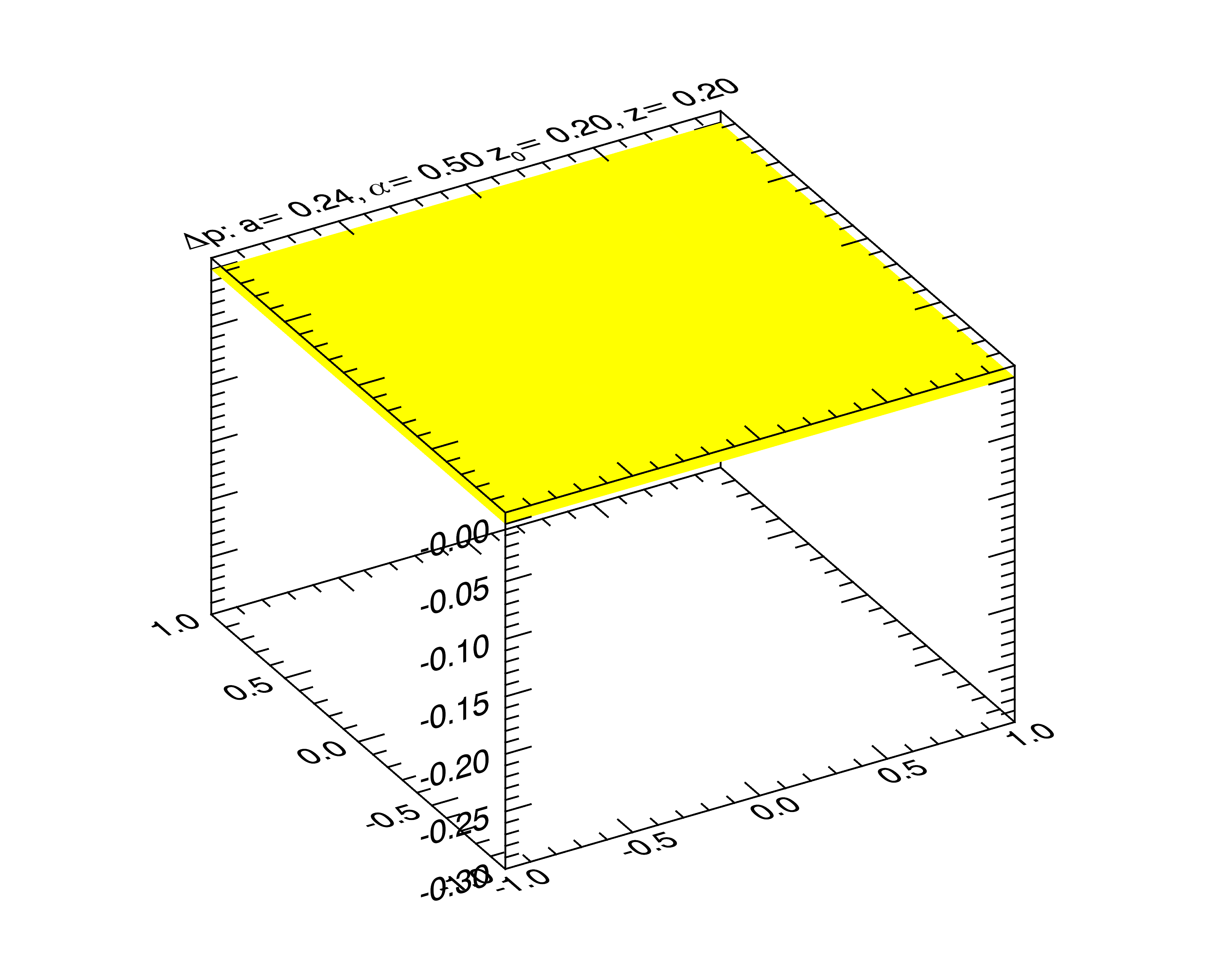}
 \hspace*{-0.03\textwidth}
\includegraphics[width=0.515\textwidth,clip=]{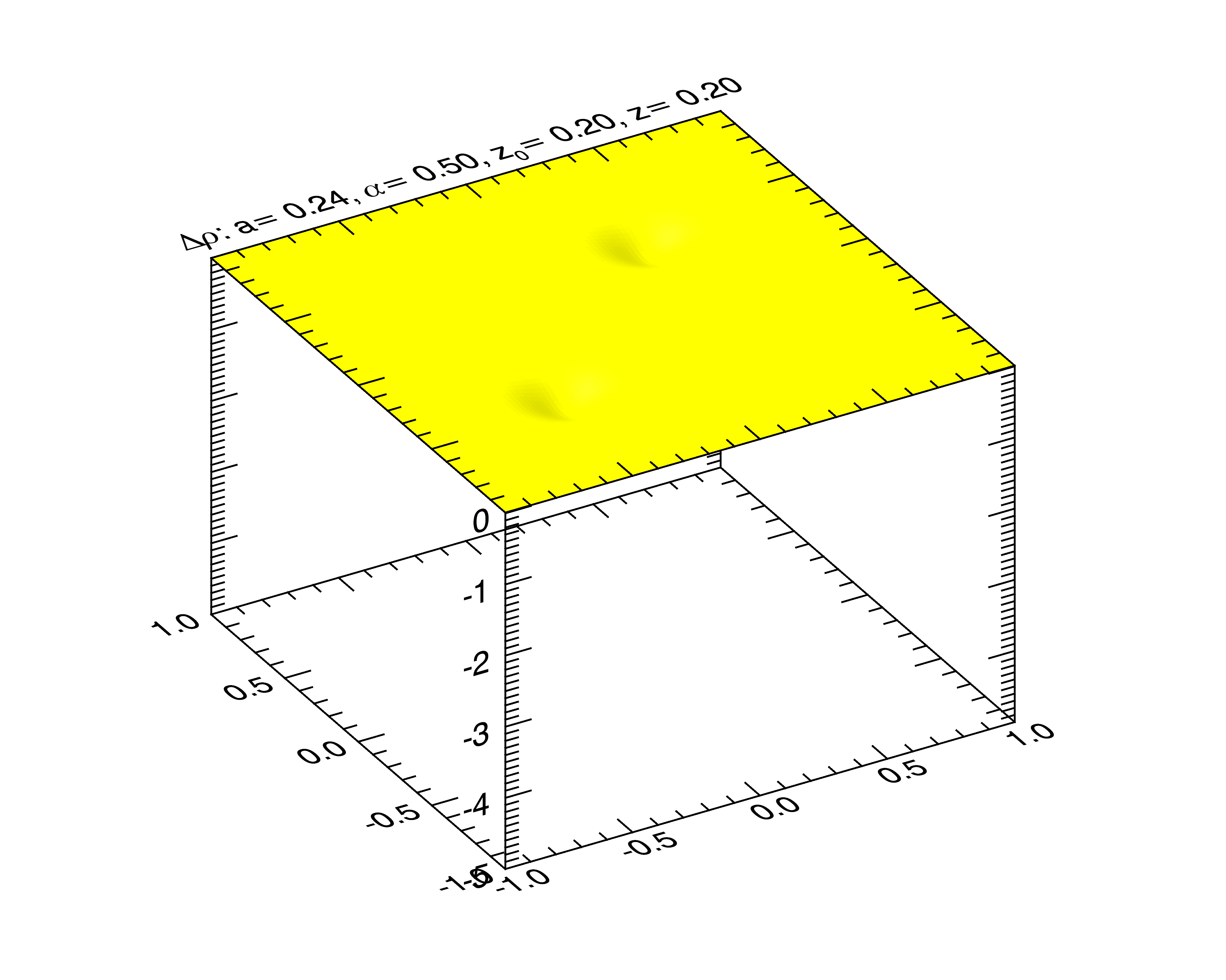}
}
 \vspace{-0.35\textwidth}   
     \centerline{\large \bf     
      \hspace{0.0 \textwidth}  \color{black}{(e)}
      \hspace{0.415\textwidth}  \color{black}{(f)}
         \hfill}
     \vspace{0.31\textwidth}    

 \caption{Surface plots of $\Delta p$ and $\Delta \rho$ at the heights $z=0$ ((a) and (b)), $z=z_0/2$ 
 ((c) and (d))and $z=z_0$ ((e) and (f)), showing the
 variation of pressure and density in the $x$- and $y$-directions.
}
   \label{fig:pandrho-surfaces}
\end{figure}

In Figures \ref{fig:pandrho-lineplots} and \ref{fig:pandrho-surfaces} we show the spatial variation of
\begin{equation}
\Delta p = - f(z) \frac{B_z^2}{2 \mu_0}  
\label{eq:def-delp}
\end{equation}
and
\begin{equation}
\Delta \rho =\frac{1}{g}\left( \deriv{f}{z} \frac{B_z^2}{2 \mu_0}+\frac{f}{\mu_0} \bb\cdot \grad B_z \right).
\label{eq:def-delrho}
\end{equation}
In these two figures $\Delta p$ is normalised by $B_{\mathrm{max}}^2/\mu_0$, where $B_{\mathrm{max}}$ is the maximum value
of $B_z$ at $z=0$, and $\Delta \rho$ is normalised by $B_{\mathrm{max}}^2/(\mu_0 g L)$
Figure \ref{fig:pandrho-lineplots} shows the variation of $\Delta p$ and $\Delta \rho$ with height for
$0 \le z \le 2 z_0$ at the position of the maximum of $|B_z|$ on the lower boundary, {\it i.e.}\ 
$x=y=\tilde{\mu}_{x1}  \approx -0.382$. We have chosen these $x$ and $y$-coordinates because one
expects the largest variations of $\Delta p$ and $\Delta \rho$ to happen there. As one can see
both $\Delta p$ and $\Delta \rho$ have negative values and they generally increase with $z$ from
their lowest value at $z=0$ until approaching zero around $z=z_0$, as expected. 
The increase of the amplitude of $\Delta p$ and $\Delta \rho$ and their slower increase with $z$ 
with increasing value of $a$ is also obvious. The latter is caused by the Fourier modes, in particular
the lowest order mode, to decrease less fast with height when $a$ increases. 
Although a detailed analysis is a bit more complicated, one can roughly
regard the $z$ dependence of each mode as being given by $\exp(-2 \gamma z)$
below and $\exp(-2 \delta z)$ above $z\approx z_0$. As $a \to a_{\mathrm{max}}$ we have $\gamma \to 0$ for the
lowest order mode and hence the mode decreases more slowly with $z$.

We also note that $\Delta \rho$ has a local maximum and minimum just below $z_0$. This is caused
by the $\deriv{f}{z}$ term in $\Delta \rho$, which is largest at $z_0$. This term becomes the larger the
sharper the gradient of $f$ at $z=z_0$ is made, {\it i.e.}\ the smaller $\Delta z$ becomes (the
derivative actually
tends to a $\delta$-function in the limit $\Delta z \to 0$ in which case the $\tanh$-function tends
to a step function). For this reason one should not attempt to make $\Delta z$ too small.

\begin{figure}
\centerline{\hspace*{0.025\textwidth}
\includegraphics[width=0.7\textwidth,clip=]{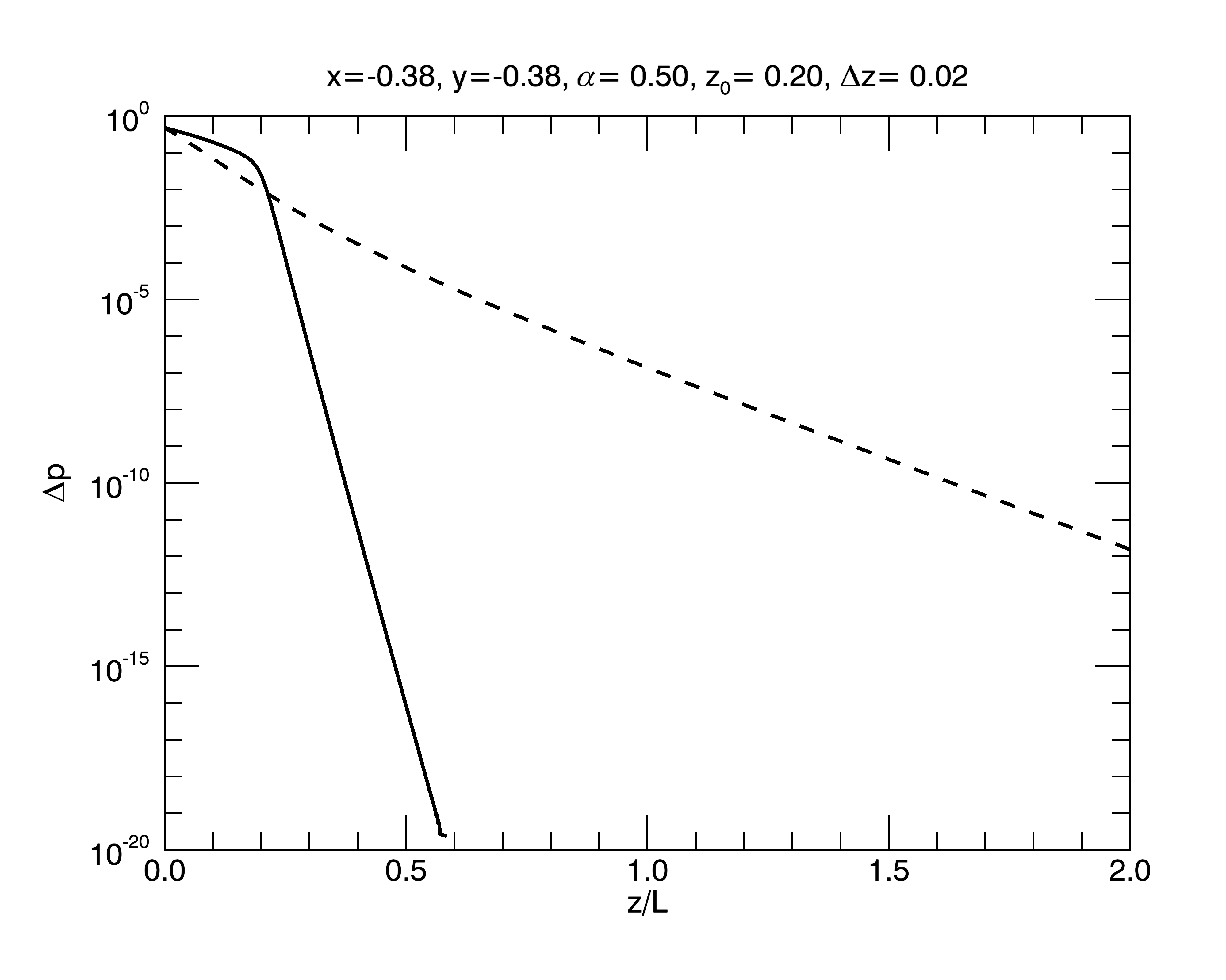}
}
 \vspace{-0.5\textwidth}   
     \centerline{\large \bf     
      \hspace{0.1\textwidth}  \color{black}{(a)}
         \hfill}
     \vspace{0.5\textwidth}    
 \hspace*{-0.03\textwidth}
\centerline{\hspace*{0.025\textwidth}
\includegraphics[width=0.7\textwidth,clip=]{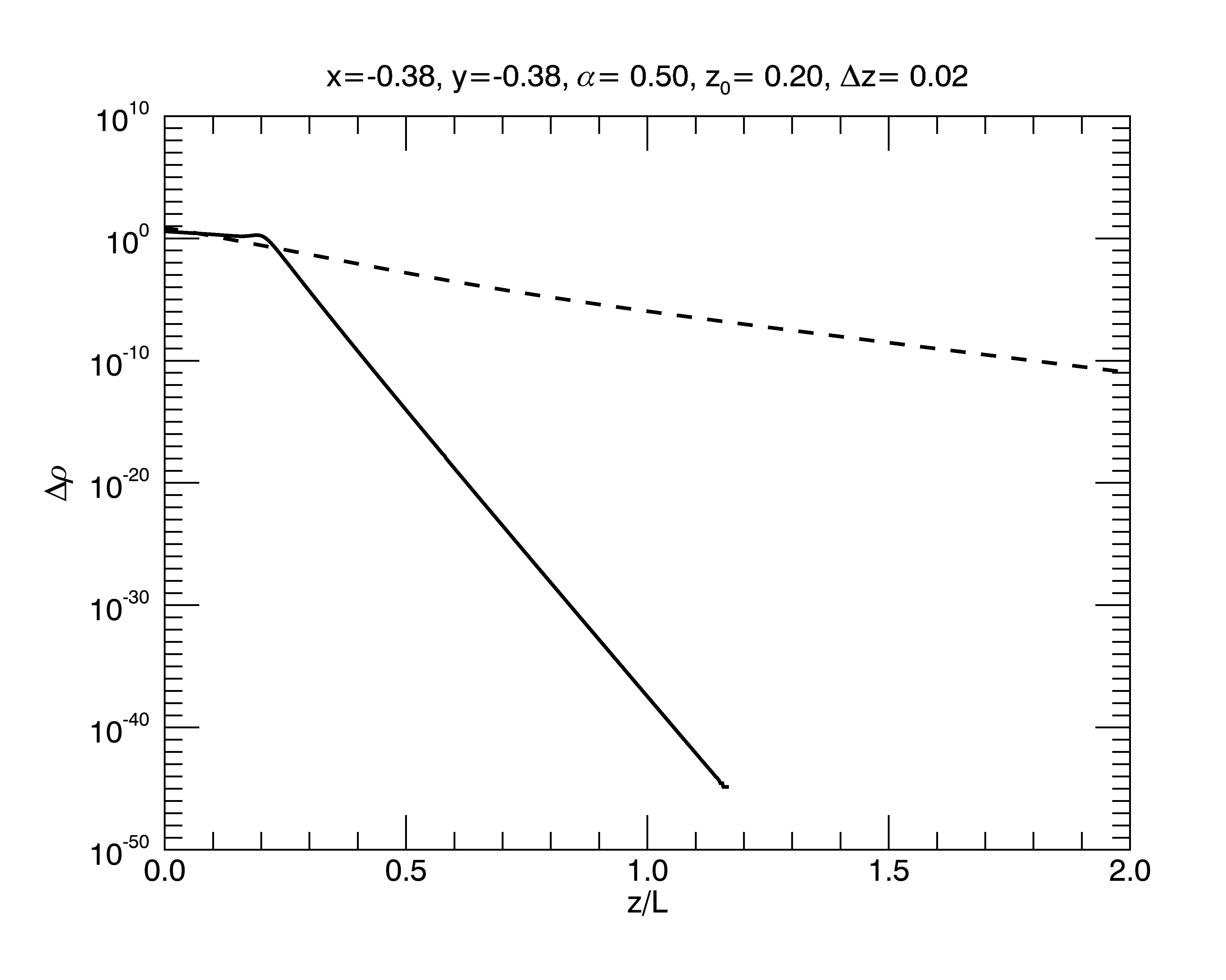}
}

 \vspace{-0.5\textwidth}   
     \centerline{\large \bf     
      \hspace{0.1\textwidth}  \color{black}{(b)}
         \hfill}
     \vspace{0.5\textwidth}    
   \caption{Comparison of the variation with height $z$ of the absolute values of the pressure (panel (a)) and density (panel(b)) deviation from a stratified background 
   atmosphere for the solutions presented in this paper (solid lines) and
   for a solution with an exponential $f(z)$ \citep{low91}. The plots have a logarithmic $y$-axis to emphasize 
   the difference of the quantities of the solutions,  especially the much faster drop-off of the solutions presented in this paper 
   above the transition height $z_0$. The parameters chosen for this plot are $\alpha = 0.5$, $a = 0.48$, $z_0/L = 0.2$ and $\Delta z = 0.1 z_0$ for the solid line, whereas we
   have chosen $\bar{a} = 0.96$ and $ \kappa = 1/z_0$ (see main text for definitions).
   The $x$- and $y$-coordinates of these plots are the position of the maximum value of $|B_z|$ on the lower boundary, 
   $x/L=\mu_x/\pi \approx -0.38$, $y/L=\mu_y/\pi \approx -0.38$.}
   \label{fig:pandrho-comparison}
\end{figure}

To keep the pressure and density positive everywhere we need to make the background pressure $p_0(z)$
and the corresponding background density $\rho_0(z) = - \frac{1}{g} \deriv{p_0}{z}$ positive and larger at every
height $z$ than the minimum values of $\Delta p$ and $\Delta \rho$ for all $x$, $y$ values at the same height
\citep[see \textit{e.g.}][]{wiegelmann:etal15}. As one can see from Figures \ref{fig:pandrho-lineplots} and \ref{fig:pandrho-surfaces}
for this new solution family this condition is only needed to be satisfied for $z < z_0$ in the case
$b=1.0$ because the contributions to the pressure and the density by $\Delta p$ and $\Delta \rho$ become
arbitrarily small very fast for $z > z_0$. 
In Figure\ \ref{fig:pandrho-comparison} we show a comparison of variation with $z$ of the absolute values of $\Delta p$ and $\Delta \rho$ (again at $x=y=\tilde{\mu}_{x1}  \approx -0.382$),
for a solution of the family presented in this paper and a solution for $f(z) = \bar{a} \exp(- \kappa z)$ \citep{low91}. We remark that in order to start with the same pressure at $z=0$ one has to set
$\bar{a} = 2 a$. We have also chosen the inverse length scale $\kappa = 1/z_0$, corresponding to the height of the transition of the solutions presented in this paper from non-force-free to force-free.
These plots clearly show that while for both sets of solutions the pressure and density deviations decrease rapidly for $z > z_0$, the decrease is much more rapid for our solutions, which, as already stated above,
reduces the need for artificially adjusting the stratified background atmosphere in this region to avoid negative density or pressure values.

One could of course argue that by increasing $\kappa$, {\it i.e.}\ reducing the length scale over which the exponential $f(z)$ and hence the pressure and density deviations decay, one could in principle achieve a similar effect for the
exponential $f(z)$. However, for this $f(z)$ this would also decrease the effect of the non-force-free part of the current density in the regions below $z_0$ and hence reduce the effect of the non-force-free part of
the current density, which somewhat defeats the purpose of using such a magnetic field model for extrapolation in the first place. Because the solutions presented in this paper can guarantee a very rapid
transition to a linear force-free magnetic field, but still
allow us to control the region $z <z_0$ separately this problem does not arise in the new family of solutions.

\section{Discussion and Conclusions}
\label{sec:discussion}

We have presented a new family of three-dimensional MHS solutions based on the general theory originally
developed by Low and co-workers \citep[\textit{e.g.}][]{low85, bogdan:low86, low91, low92}, although we have
used the alternative mathematical formulation by \citet{neukirch:rastatter99}. 
{The main motivation for trying to find these solutions was that they are of potential importance for analytical non-force-free magnetic field extrapolation methods.}
The new family of solutions
allows for the MHS nature of the equilibria to be limited to a domain below a specific pre-determined height 
with a smooth transition to a potential or linear force-free solution possible above height. 
This is achieved by choosing one of the fundamental free functions of the theory in the form of a hyperbolic tangent.
It is important
to emphasize that this is just one possibility of choosing the parameters for this equilibrium family and that
MHS solutions in the full domain are also possible.

While such a transition is also possible with the free function mentioned above chosen
in the form a decaying exponential function \citep[\textit{e.g.}][]{low91, low92, aulanier:etal98b, aulanier:etal99,wiegelmann:etal15, wiegelmann:etal17},
the new solution family allows much more control over the transition from the MHS domain to the non-MHS domain, in particular regarding the 
departures of the plasma pressure and density from a stratified atmosphere.
This property can be of importance for keeping the plasma pressure and density of the solution positive and should make
this equilibrium family interesting for magnetic field extrapolation purposes when a non-force-free layer has to be included. 
We emphasize that  we do not regard these analytical extrapolation methods as replacements for numerical methods for non-force-free magnetic field extrapolation, but
as a numerically relatively cheap complementary method, which could be used as an initial "quick look" tool. Obviously, the limitations that arise from having to make
a number of strong assumptions to allow analytical progress have to born in mind when applying these methods.

%

%

%
 \appendix   

\section{Fourier Decomposition of Example Magnetic Field}

\label{app:fourier}

Using the identity \citep[\textit{e.g.}][]{abramowitz:stegun65,Olver:2010:NHMF}
\begin{equation}
\exp[x \cos(y)] = \mbox{I}_0(x) + 2 \sum\limits_{n=1}^\infty \mbox{I}_n(x) \cos(n y),
\label{eq:expcos-in-I}
\end{equation}
plus trigonometric identities, it is straightforward (albeit a bit tedious) to write the expression for $B_z(x,y,0)$ in Equation \ref{eq:magnetogram}
in the form
\begin{eqnarray}
B_z(x,y,0) &=& \sum\limits_{n=1}^\infty  \sum\limits_{m=1}^\infty [a_{nm} \sin(n \bar{x}) \sin(m \bar{y})  + b_{nm}  \sin(n \bar{x}) \cos(m \bar{y}) + \nonumber\\
                  & & \mbox{\hspace{1.5cm}}                                          c_{nm}\cos(n \bar{x}) \sin(m \bar{y}) + d_{nm}  \cos(n \bar{x}) \cos(m \bar{y}) + \nonumber\\
                  & &     \sum\limits_{n=1}^\infty [b_{n0} \sin(n \bar{x}) + d_{n0}  \cos(n \bar{x}) ] +                                                                                     \nonumber\\
                  & &      \sum\limits_{m=1}^\infty  [c_{0m} \sin(m \bar{y}) + d_{0m}  \cos(m \bar{y})   ]   ,        
\label{eq:Bz_fourierexp}                               
\end{eqnarray}
with
\begin{eqnarray}
a_{nm}  &=& \frac{B_0}{\pi^2} \left[I_{f1,nm} \sin(n \bar{\mu}_{x1})  \sin(m \bar{\mu}_{y1}) -   I_{f2,nm} \sin(n \bar{\mu}_{x2})  \sin(m \bar{\mu}_{y2}) \right]  ,      \label{eq:anm-def} \\
b_{nm}  &=& \frac{B_0}{\pi^2} \left[I_{f1,nm} \sin(n \bar{\mu}_{x1})  \cos(m \bar{\mu}_{y1}) -   I_{f2,nm} \sin(n \bar{\mu}_{x2})  \cos(m \bar{\mu}_{y2}) \right],         \label{eq:bnm-def} \\
c_{nm}  &=& \frac{B_0}{\pi^2} \left[I_{f1,nm} \cos(n \bar{\mu}_{x1})  \sin(m \bar{\mu}_{y1}) -   I_{f2,nm} \cos(n \bar{\mu}_{x2})  \sin(m \bar{\mu}_{y2}) \right] ,        \label{eq:cnm-def} \\
d_{nm}  &=& \frac{B_0}{\pi^2} \left[I_{f1,nm} \cos(n \bar{\mu}_{x1})  \cos(m \bar{\mu}_{y1}) -   I_{f2,nm} \cos(n \bar{\mu}_{x2})  \cos(m \bar{\mu}_{y2}) \right]  ,       \label{eq:dnm-def}
\end{eqnarray}
where the factors $I_{fi,nm}$ are defined as
\begin{equation}
I_{fi,nm} = \frac{ \mbox{I}_n(\tilde{\kappa}_{xi} ) \mbox{I}_m(\tilde{\kappa}_{yi}) }{ 2 \mbox{I}_0(\tilde{\kappa}_{xi}) \mbox{I}_0(\tilde{\kappa}_{yi})}  \times
\left\{  \begin{array}{ll} 
                                  1 & \mbox{for $m=0$, $n>0$ or $n=0$, $m>0$  } \\
                                  2 & \mbox{for $m>0$ and $n>0$} 
                                   \end{array} \right. .
\label{eq:I-f-def}
\end{equation}
The corresponding series for $B_z$ is given by
\begin{eqnarray}
B_z(x,y,0) &=& \sum\limits_{n=1}^\infty  \sum\limits_{m=1}^\infty k_{nm}^2 \bar{\Phi}_{nm}(0)[\bar{a}_{nm} \sin(n \bar{x}) \sin(m \bar{y})  + \bar{b}_{nm}  \sin(n \bar{x}) \cos(m \bar{y}) + \nonumber\\
                  & & \mbox{\hspace{1.5cm}}                                          \bar{c}_{nm}\cos(n \bar{x}) \sin(m \bar{y}) + \bar{d}_{nm}  \cos(n \bar{x}) \cos(m \bar{y}) ]+ \nonumber\\
                  & &     \sum\limits_{n=1}^\infty  k_{n0}^2 \bar{\Phi}_{nm}(0) [\bar{b}_{n0} \sin(n \bar{x}) + \bar{d}_{n0}  \cos(n \bar{x}) ] +                                                                                     \nonumber\\
                  & &      \sum\limits_{m=1}^\infty  k_{0m}^2 \bar{\Phi}_{nm}(0) [\bar{c}_{0m} \sin(m \bar{y}) + \bar{d}_{0m}  \cos(m \bar{y})   ]   ,        
\label{eq:Bz_fourierexp-sol}                               
\end{eqnarray}
with $\bar{\Phi}_{nm}(z)$ the solution of Equation \ref{eq:Phibar-eckart-equation} for the wave vector
\begin{equation}
k^2_{nm} = \frac{\pi^2}{L^2}(n^2 + m^2) .
\end{equation}
We remark that $n$ or $m$ are allowed to take on the value $0$, but not both at the same time. Comparing Equations \ref{eq:Bz_fourierexp} and \ref{eq:Bz_fourierexp-sol} one can easily see that
$\bar{a}_{nm} = a_{nm}/ (k_{nm}^2 \bar{\Phi}_{nm}(0))$ {\it etc}.

%
 \begin{acks}
 For producing some of the figures in this paper the authors have used of the IDL routines for hypergeometric functions by Michele Cappellari
 (\burl{https://github.com/surftour/astrotools/blob/master/idlstuff/IDL\_kin/jam\_modelling/utils/hypergeometric2f1.pro} and\\
  \burl{http://www-astro.physics.ox.ac.uk/\~mxc/software/}).
 TN acknowledges financial support by the UK's Science and Technology Facilities Council (STFC), Consolidated Grants ST/K000950/1, ST/N000609/1 and ST/S000402/1
 and would like to thank the colleagues at the Max-Planck-Institute for Solar System Research for their hospitality during his visits in 2014 and 2015 when the 
 idea for this research originated.
TW acknowledges financial support by DFG Grant WI 3211/4-1.
 \end{acks}

\begin{acks}[Disclosure of Potential Conflicts of Interest]

The authors declare that they have no conflicts of interest.

 \end{acks}

%
%
%

\end{article} 
\end{document}